\documentclass[review]{elsarticle}

\usepackage{lineno,hyperref,amsmath}
\modulolinenumbers[5]

\journal{Nuc. Instr. Meth. A}

\begin{document}

\begin{frontmatter}

\title{The bias of the unbiased estimator: a study of the iterative application of the BLUE method.}

\author[2]{Luca Lista}
\address[2]{INFN Sezione di Napoli}

\begin{abstract}
The best linear unbiased estimator (BLUE) is a popular statistical method adopted to combine multiple
measurements of the same observable taking into account individual
uncertainties and their correlations. The method is unbiased by construction
if the true uncertainties and their correlations are known, but it
may exhibit a bias if uncertainty estimates are used in place of the true ones,
in particular if those estimated uncertainties depend on measured values.
This is the case for instance when contributions to the total 
uncertainty  are known as relative uncertainties.
In those cases, an iterative application of the BLUE method may reduce the
bias of the combined measurement.
The impact of the iterative approach compared to the standard BLUE application is 
studied for a wide range of possible values of uncertainties and their correlation in 
the case of the combination of two measurements.
\end{abstract}


\end{frontmatter}

\linenumbers

\section{Introduction}
The application of the best linear unbiased estimator (BLUE) method
to combine correlated estimates of a single physical quantity
is due to L. Lyons {\it et al.}~\cite{BLUE_ref1}. Assuming to have two or more
measurements $x_1 \pm \sigma_1,\,\cdots,\,x_n\pm \sigma_n$ of the same observable $x$,
knowing their Gaussian uncertainties and their correlations, a generic linear estimator of $x$ can be written as:
\begin{equation}
\hat{x} = \sum_{i=1}^n x_i w_i\,.
\label{eq:blue}
\end{equation} 
The above estimator is unbiased if the sum of the weights is equal to one.
The linear unbiased estimator having the smallest variance can be determined by finding 
the weights $w_1,\, \cdots,\, w_n$ that minimize the following $\chi^2$, imposing the constraint
$\sum_i w_i = 1$:
\begin{equation}
\chi^2 = 
\left(x_1 - \hat{x}\,\cdots,\, x_n - \hat{x} \right) 
C^{-1}
\left(\begin{array}{c}
x_1 - \hat{x} \\
\cdots\\
x_n - \hat{x}
\end{array}\right)\,,
\label{eq:blue-chi2}
\end{equation}
where $C$ is the covariance matrix of the $n$ measurements.
In the following, for simplicity, the case of two measurements ($n=2$) is assumed.
The $\chi^2$ minimization from Eq.(\ref{eq:blue-chi2}) for $n=2$ gives the weights:
\begin{eqnarray}
w_1 & = & \frac{\sigma_2^2-\rho\sigma_1\sigma_2}{\sigma_1^2-2\rho\sigma_1\sigma_2+\sigma_2^2}\,, \\
w_2 & = & \frac{\sigma_1^2-\rho\sigma_1\sigma_2}{\sigma_1^2-2\rho\sigma_1\sigma_2+\sigma_2^2}\,, 
\end{eqnarray}
where $\rho$ is the correlation coefficient of the uncertainties affecting both measurements $x_1 \pm \sigma_1$ 
and $x_2 \pm \sigma_2$  (e.g. systematic uncertainties can be correlated across measurements, 
or luminosity uncertainty may affect different cross section measurements).
The uncertainty of the combined value $\hat{x}$ can be determined as
the standard deviation of the BLUE estimator, which for a Gaussian distribution is:
\begin{equation}
\sigma_{\hat{x}} = \sqrt{
\frac{\sigma_1\sigma_2(1-\rho^2)}{\sigma_1^2-2\rho\sigma_1\sigma_2+\sigma_2^2}
}\,.
\label{eq:blueerr}
\end{equation}

L. Lyons {\it et al.} remarked the limitation of the application of the BLUE method in 
the combination of lifetime measurements where uncertainty estimates $\hat{\sigma_i}$ of the
true unknown uncertainties $\sigma_i$ were used, and those estimates
had a dependency on the measured lifetime.
This issue was addressed in a later paper~\cite{LyonsMartinSaxon},
which also demonstrated that the application
of the BLUE method violates, in that case, the ``combination principle'':
if the set of measurements is split into a number of subsets,
then the combination is first performed in each subset and finally all
subset combinations are combined into a single combination, this result
differs from the combination of all individual results of the entire set.

For this case, Ref.~\cite{LyonsMartinSaxon} recommended to apply iteratively 
the BLUE method, rescaling at each iteration the uncertainty estimates
according to the central value obtained with the BLUE method in the
previous iteration, until the sequence converges to a stable result. In this
way the bias of the BLUE estimate is reduced compared
to the application of the BLUE method with no iteration (in the following
this original application of the method is referred to as ``standard'' BLUE method).
Also, the ``combination principle'' is respected to a good approximation level, at least for the 
mentioned B-meson lifetime study, 
in the sense that the combination of partial combinations is very close to the
combination of all available individual measurements.


One may wonder how those conclusions may be valid in general.
The presented study  attempts to give an answer exploring a 
wide range of possible uncertainty values and their correlations 
for the combination of two measurement.

\section{Applying the BLUE method iteratively}

The estimates of uncertainties and their correlation are assumed to be known as a function of the measured
values of the true quantity $x$. Given the measured values $x_1$ and $x_2$, 
the uncertainties and their correlation can be written as:
\begin{equation}
\begin{array}{lcl}
\hat{\sigma}_1 & = & \hat{\sigma}_1(x_1)\,, \\
\hat{\sigma}_2 & = & \hat{\sigma}_2(x_2)\,, \\
\hat{\rho} & = & \hat{\rho}(x_1, x_2)\,.
\end{array}\nonumber
\end{equation}
The application of the standard BLUE method including the estimated uncertainties
in Eq.~(\ref{eq:blue}) gives the combined value
\begin{equation}
\hat{x} = \frac{(\hat{\sigma}_2^2-\hat{\rho}\hat{\sigma}_1\hat{\sigma}_2)x_1 + 
(\hat{\sigma}_1^2-\hat{\rho}\hat{\sigma}_1\hat{\sigma}_2) x_2}{\hat{\sigma}_1^2-2\hat{\rho}\hat{\sigma}_1\hat{\sigma}_2+\hat{\sigma}_2^2}\,.
\label{eq:blue2}
\end{equation}
Since $\hat{\sigma}_1$, $\hat{\sigma}_2$ and $\hat{\rho}$ are not the true uncertainties and correlation,
but their estimates, it is not guaranteed that $\hat{x}$ is unbiased and that it has the smallest possible (``best'')
variance. Indeed, in many possible cases $\hat{x}$ exhibits a bias, as  will be shown in the following.

A classic example of this effect is the combination of two measurements whose uncertainty estimates are
proportional to the square root of the measured values, as typically from a Poissonian event counting.
Let's consider two uncorrelated measurements of the expected yield in a 
Poissonian counting experiment:
\begin{eqnarray}
\hat{n}_1 & = & n_1 \pm \sqrt{n_1}\,, \\
\hat{n}_2 & = & n_2 \pm \sqrt{n_2} \,.
\end{eqnarray}
The maximum-likelihood combination of the two measurement, which in this case is unbiased, is: 
\begin{equation}
\hat{n}_{\textrm{max-lik.}}= \frac{1}{2}(n_1+n_2 \pm \sqrt{n_1+n_2})\,.
\label{eq:poiscomb}
\end{equation}
Since the two measurements are uncorrelated the BLUE estimate is a
weighted average with weights $w_i \propto 1/n_i$, $i=1$, $2$, which in this case results in 
the harmonic average:
\begin{equation}
\hat{n}_{\mathrm{BLUE}}= \frac{2 n_1 n_2}{n_1+n_2} \pm \sqrt{\frac{n_1 n_2}{n_1+n_2} } \,.
\label{eq:poisblue}
\end{equation}
Compared to Eq.~(\ref{eq:poiscomb}), Eq.~(\ref{eq:poisblue}) exhibits a bias induced by the fact that
a measurement with a downward fluctuation achieves a larger weight pulling down the combination,
while the corresponding effect of an upward fluctuation is reduced due to the specific dependence of uncertainties
on the measured values.

Since $\hat{x}$ is a ``better'' estimate of $x$ than the individual measurements $x_1$ and $x_2$, one may recompute 
uncertainties and their correlation at the new combined value $\hat{x}^{(1)}=\hat{x}$ and obtain new estimates for
$\hat{\sigma}_1$, $\hat{\sigma}_2$ and $\hat{\rho}$:
\begin{equation}
\begin{array}{lcl}
\hat{\sigma}_1^{(1)} & = & \hat{\sigma}_1(\hat{x}^{(1)})\,, \\
\hat{\sigma}_2^{(1)} & = & \hat{\sigma}_2(\hat{x}^{(1)})\,, \\
\hat{\rho}^{(1)} & = & \hat{\rho}(\hat{x}^{(1)}, \hat{x}^{(1)})\,.
\end{array}\nonumber
\end{equation}
The BLUE method can be applied again using the new uncertainty estimates
$\hat{\sigma}_1^{(1)}$, $\hat{\sigma}_2^{(1)}$  and their correlation estimate $\hat{\rho}^{(1)}$,
and a new central value estimate $\hat{x}^{(2)}$ can be obtained. Uncertainties
and their correlation can be recomputed once again:
\begin{equation}
\begin{array}{lcl}
\hat{\sigma}_1^{(2)} & = & \hat{\sigma}_1(\hat{x}^{(2)})\,, \\
\hat{\sigma}_2^{(2)} & = & \hat{\sigma}_2(\hat{x}^{(2)})\,, \\
\hat{\rho}^{(2)} & = & \hat{\rho}(\hat{x}^{(2)}, \hat{x}^{(2)})\,,
\end{array}\nonumber
\end{equation}
and the method can be applied iteratively,
until it converges. If the sequence $\hat{x}^{(1)}, \cdots, \hat{x}^{(n)}$ converges, its limit
$\hat{x}_{\mathrm{it}}$ satisfies the following condition:
\begin{equation}
\hat{x}_{\mathrm{it}} = \frac{
(\hat{\sigma}_2(\hat{x}_{\mathrm{it}})^2 - \hat{\rho}(\hat{x}_{\mathrm{it}},\hat{x}_{\mathrm{it}})\hat{\sigma}_1(\hat{x}_{\mathrm{it}})\hat{\sigma}_2(\hat{x}_{\mathrm{it}})) x_1  + 
(\hat{\sigma}_1(\hat{x}_{\mathrm{it}})^2-\hat{\rho}(\hat{x}_{\mathrm{it}},\hat{x}_{\mathrm{it}})\hat{\sigma}_1(\hat{x}_{\mathrm{it}})\hat{\sigma}_2(\hat{x}_{\mathrm{it}})) x_2
}{\hat{\sigma}_1(\hat{x}_{\mathrm{it}})^2-2\hat{\rho}(\hat{x}_{\mathrm{it}},\hat{x}_{\mathrm{it}})\hat{\sigma}_1(\hat{x}_{\mathrm{it}})\hat{\sigma}_2(\hat{x}_{\mathrm{it}})+\hat{\sigma}_2(\hat{x}_{\mathrm{it}})^2}\,.
\label{eq:iterblue}
\end{equation}
An estimate of the variance of $\hat{x}_{\mathrm{it}}$ can be determined from Eq.~(\ref{eq:blueerr}) using individual
uncertainty values and their correlation evaluated at $\hat{x}_{\mathrm{it}}$. This estimate, anyway, reproduces the true standard deviation
of the estimator's distribution if the BLUE hypotheses are fulfilled, which is not necessarily the case with the presented assumptions,
and deviations of this error estimate from the true standard deviation may occur, as will be discussed in the following.

One may argue whether $\hat{x}_{\mathrm{it}}$ has better statistical properties
than $\hat{x}$, in particular whether $\hat{x}_{\mathrm{it}}$ has a smaller bias than $\hat{x}$.
In order to simplify the problem of studying any possible
dependence of $\hat{\sigma}_1$, $\hat{\sigma}_2$ and $\hat{\rho}$ on $x$, uncertainties squared are assumed to be
the sum in quadrature of a constant term plus a term that depends linearly on the corresponding measured value:
\begin{eqnarray}
\hat{\sigma}_1(x_1)^2 & = & \sigma_1^2 + (r_1 x_1)^2 \label{eq:sigmalin1}\,, \\
\hat{\sigma}_2(x_2)^2 & = & \sigma_2^2 + (r_2 x_2)^2 \label{eq:sigmalin2}\,.
\end{eqnarray}
This is the case when combining cross-section measurements where contributions to the uncertainty
are due to acceptance, efficiencies and integrated luminosity which propagate into relative uncertainties on the measured cross section.

Let's assume $\rho_0$ to be the correlation between the uncertainty contributions $\sigma_1$ and $\sigma_2$
and $\rho_{\mathrm{r}}$ to be the correlation of the uncertainty contributions $r_1x_1$ and $r_2x_2$; moreover,
let's assume that $\rho_0$ and $\rho_{\mathrm{r}}$ do not depend on the measured values $x_1$ and $x_2$. In this case
the estimated covariance matrix of the two measurements is:
\begin{equation}
\hat{C} = \left(
\begin{array}{cc}
 \sigma_1^2 + (r_1 x_1)^2  &  \rho_0\sigma_1\sigma_2 + \rho_{\mathrm{r}} r_1 r_2 x_1 x_2 \\
  \rho_0\sigma_1\sigma_2 + \rho_{\mathrm{r}} r_1 r_2 x_1 x_2 & \sigma_2^2 + (r_2 x_2)^2
\end{array}
\right)\label{eq:bluecovmtx}
\end{equation}
and the overall correlation of $\hat{\sigma}_1(x_1)$ and $\hat{\sigma}_2(x_2)$ is given by:
\begin{equation}
\hat{\rho}(x_1, x_2) = \frac{\rho_0\sigma_1\sigma_2 + \rho_{\mathrm{r}} r_1 r_2 x_1 x_2}{\sqrt{(\sigma_1^2 + r_1^2 x_1^2)(\sigma_2^2 + r_2^2 x_2^2)}}\,.\label{eq:sigmalincov}
\end{equation}

\subsection{Special cases}

In the special case in which $r_1 = r_2 = 0$, uncertainty and correlation estimates do not depend on the measured values of $x$:
$\hat{\sigma}_1(x_1) = \sigma_1$, $\hat{\sigma}_2(x_2)=\sigma_2$, $\hat{\rho}(x_1, x_2) = \rho_0$.
Assuming those estimates to be unbiased, they must coincide with the true value. In this case, the iterative procedure 
converges at the first iteration and coincides with the result of the standard BLUE method.
Since the conditions for the validity of the standard BLUE method are fulfilled, the estimate $\hat{x} = \hat{x}_{\mathrm{it}}$ is unbiased and has the smallest possible variance.

Another special case is when $\sigma_1 = \sigma_2 = 0$, i.e. $r_1$ and $r_2$ are the {\it total} relative uncertainties:
$\hat{\sigma}_1(x_1) = r_1 x_1$, $\hat{\sigma}_2(x_2) = r_2 x_2$, $\hat{\rho}(x_1, x_2) =  \rho_{\mathrm{r}}$.
The iterative BLUE method then converges in two iterations and gives, from Eq.~(\ref{eq:iterblue}):
\begin{equation}
\hat{x}_{\mathrm{it}} = \frac{(r_2^2-\rho_{\mathrm{r}} r_1 r_2)x_1 + (r_1^2-\rho_{\mathrm{r}} r_1 r_2) x_2}{r_1^2-2\rho_{\mathrm{r}} r_1 r_2+ r_2^2}\,.
\label{eq:bluerel}
\end{equation}
The above expression is similar to the standard BLUE formula in Eq.~(\ref{eq:blue2}), but uses the relative uncertainties $r_1$ and $r_2$ instead
of the absolute ones. In this case, if one knew the true value of $x$ to be $x_0$, the standard BLUE method 
could be applied using the true uncertainties $r_1 x_0$ and $r_2 x_0$
and in Eq.~(\ref{eq:blue2}) the factor $x_0^2$ would cancel, leading to Eq.~(\ref{eq:bluerel}), which is independent on $x_0$.
In this special case the iterative application of the BLUE method would lead to the BLUE estimate
applied in the case one knew the true uncertainties.
Hence, again the iterative BLUE estimate is in this case unbiased and has the minimum variance.
Applying instead the standard BLUE method using the estimated uncertainties $\hat{\sigma}_1(x_1) = r_1 x_1$ and
$\hat{\sigma}_2(x_2) = r_2 x_1$ and their correlation $\hat{\rho}(x_1, x_2) =  \rho_{\mathrm{r}}$
would result in general in a biased estimate, hence the iterative approach provides a better estimator
than the standard one also in this case. The level of improvement in the bias gained using the iterative method depends
on the actual parameter values.

One may ask whether in the general case of Eq.~(\ref{eq:sigmalin1}) and (\ref{eq:sigmalin2}),
with $\sigma_1$, $\sigma_2$, 
$r_1$ and $r_2$ not necessarily null, the iterative method has smaller bias than the standard one
as in the two extreme special cases mentioned above.
The analytical demonstration of this statement requires non-trivial integrations. 
In the following section a parametric Monte Carlo study is applied
to address this question numerically.

\section{Study of the bias using a Monte Carlo method}

The assumed true value of $x$ is taken as $x_0 = 1$, without loss of generality.
Given the high dimensionality of the problem, 500\,000 possible sets
of the parameters $\sigma_1, \sigma_2, r_1, r_2, \rho_0$ and $\rho_{\mathrm{r}}$ are randomly chosen
using a uniform sampling limited to the ranges $\sigma_1, \sigma_2 < x_0$
and $r_1, r_2 < 1$ (100\% relative uncertainty contributions), while the entire interval $[-1,1]$ of possible values of the
correlations $\rho_0$ and $\rho_{\mathrm{r}}$ is considered. For each extracted parameter set, 500\,000 random extractions
of the measured values $x_1$ and $x_2$ are generated using a two-dimensional
Gaussian distribution having the covariance matrix from Eq.~(\ref{eq:bluecovmtx}).
The BLUE method is applied using both the standard and the iterative algorithm
and the combined values and the corresponding uncertainty estimates are determined for the
 extracted values $x_1$ and $x_2$.
The iterative BLUE method is stopped when two subsequent iterations differ less than $10^{-5}$.
The average values $\left<\hat{x}\right>$ and $\left<\hat{x}_{\mathrm{it}}\right>$ 
of the standard and the iterative BLUE estimate, respectively, 
computed on the 500\,000 extracted measurement pairs corresponding to each parameter set
are used to determine the bias of both methods. The distributions of the pulls, defined as:
\begin{eqnarray}
p  & = & \frac{\hat{x}-x_0}{\sigma_{\hat{x}}} \,, \label{eq:pull1} \\
p_{\mathrm{it}} & = & \frac{\hat{x}_{\mathrm{it}}-x_0}{\sigma_{\hat{x}_{\mathrm{it}}}}\,, \label{eq:pull2}
\end{eqnarray}
for the standard and the iterative methods, respectively, are used to determine
the standard deviation of the estimators' distributions to be compared with the BLUE uncertainty estimate
from Eq.~(\ref{eq:blueerr}). 

The properties of the standard and iterative BLUE estimators are studied in particular as a function
of the amount of relative uncertainties, $r_1$ and $r_2$, and as a function of the ratios of relative
to constant uncertainties, $r_1/\sigma_1$ and $r_2/\sigma_2$. If the ratios
$r_1/\sigma_1$ and $r_2/\sigma_2$ are small, then the hypotheses for the application of 
the standard BLUE method, that assume uncertainties independent on the measured values, is close to be fulfilled.
Hence one may expect the iterative and standard estimators should be close to each other in this case.

\section{Results}

\begin{figure}[htbp]
 \begin{center}
 \includegraphics[width=0.49\textwidth]{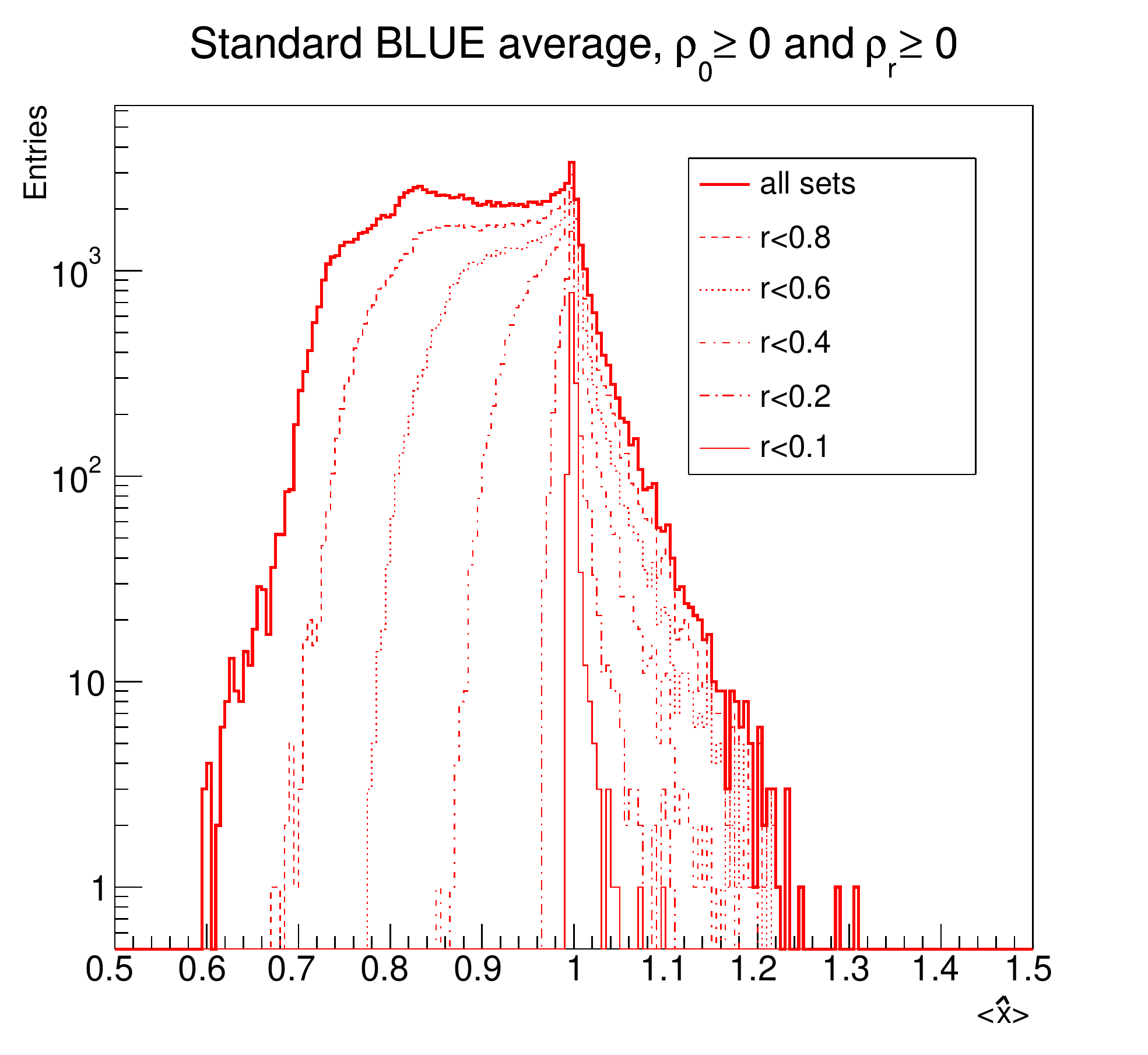}
 \includegraphics[width=0.49\textwidth]{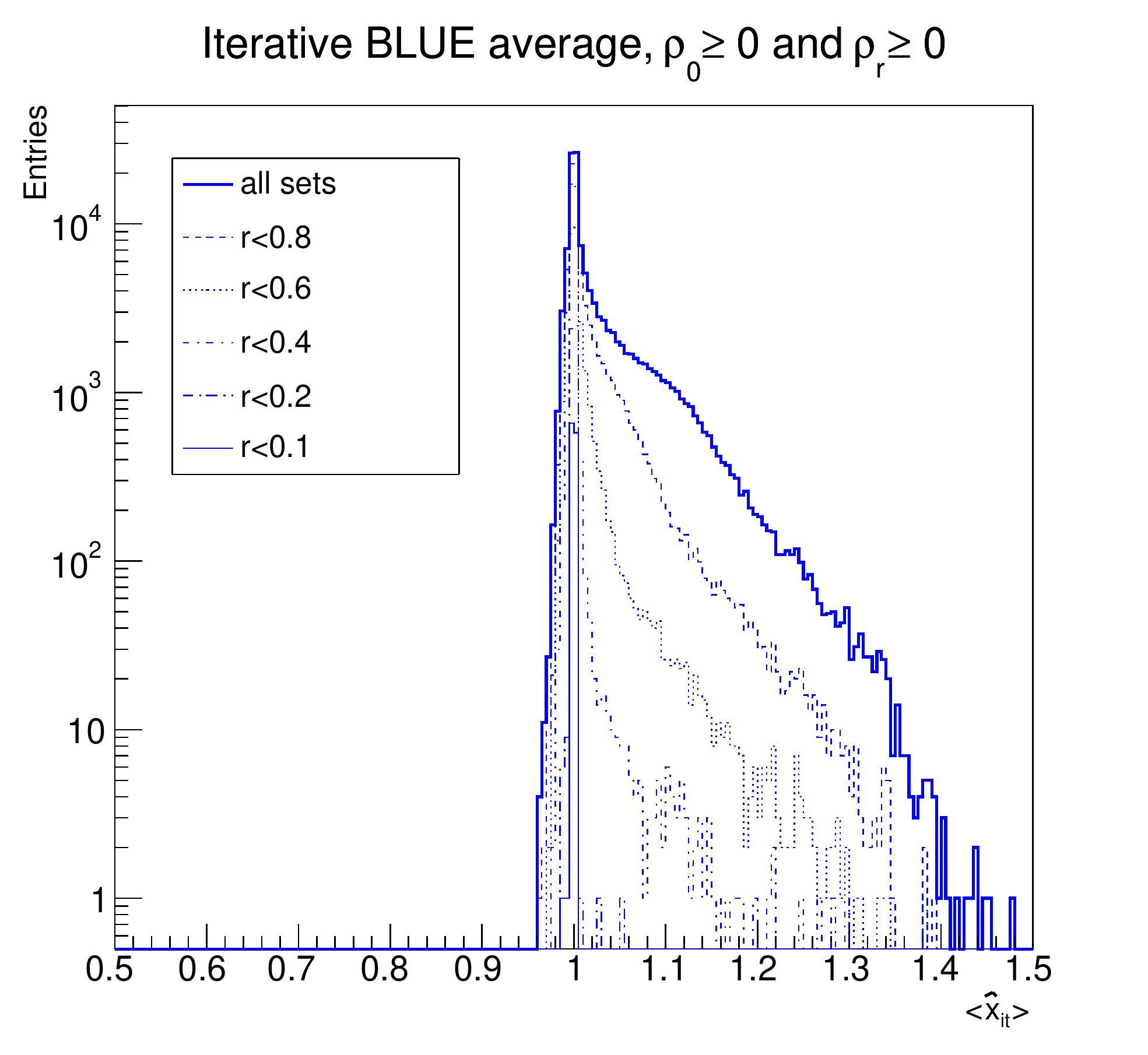}
 \includegraphics[width=0.49\textwidth]{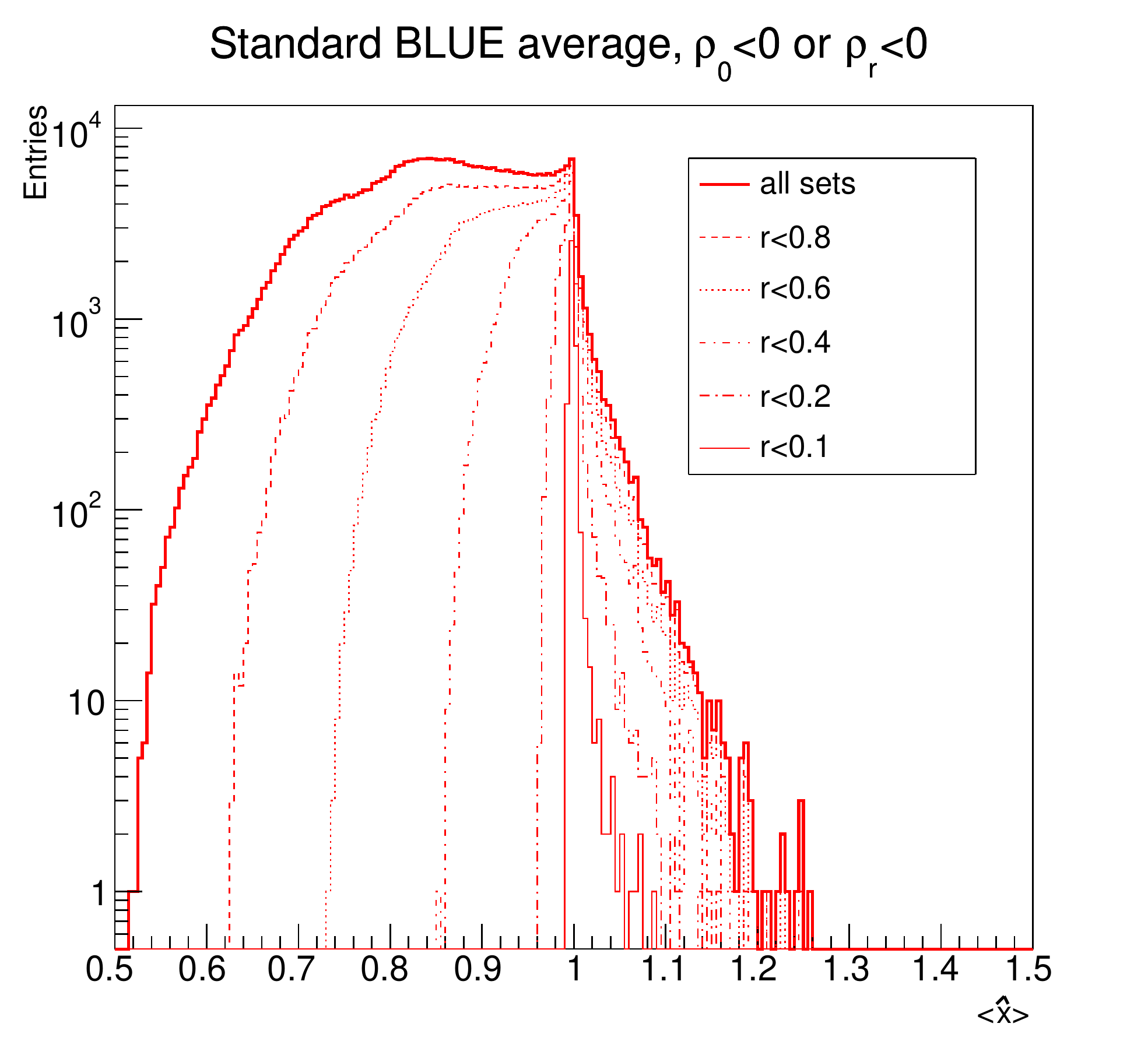}
 \includegraphics[width=0.49\textwidth]{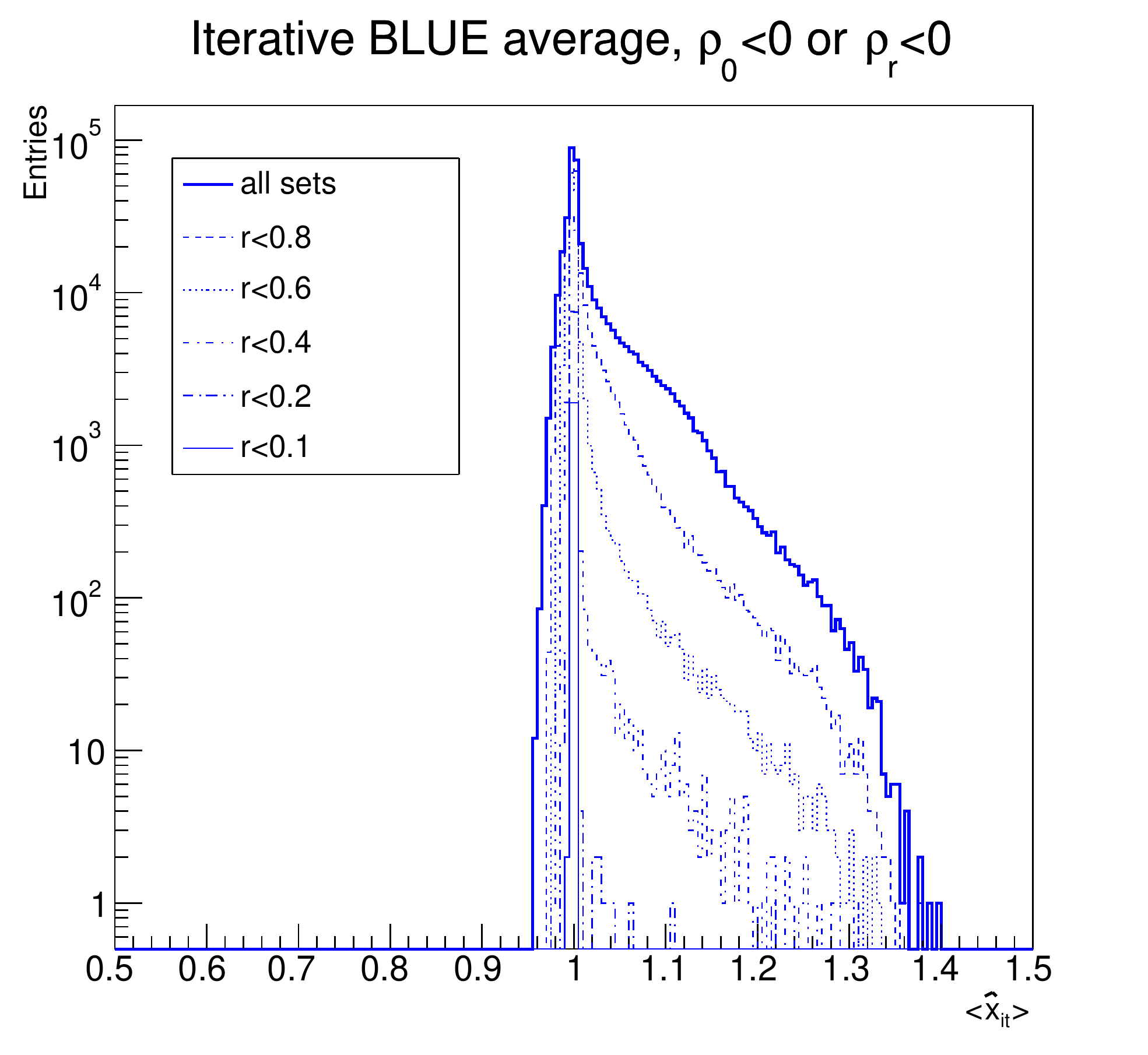}
 \caption{Distribution of the average value of the standard  (left) and iterative (right) BLUE estimates for different limits on $r$ 
 for $\rho_0\ge 0$ and $\rho_{\mathrm{r}}\ge 0$ (top) and for $\rho_0< 0$ or $\rho_{\mathrm{r}}<0$ (bottom).}
\label{fig:bias-avg1}
\end{center}
\end{figure}

\begin{figure}[htbp]
 \begin{center}
    \includegraphics[width=0.49\textwidth]{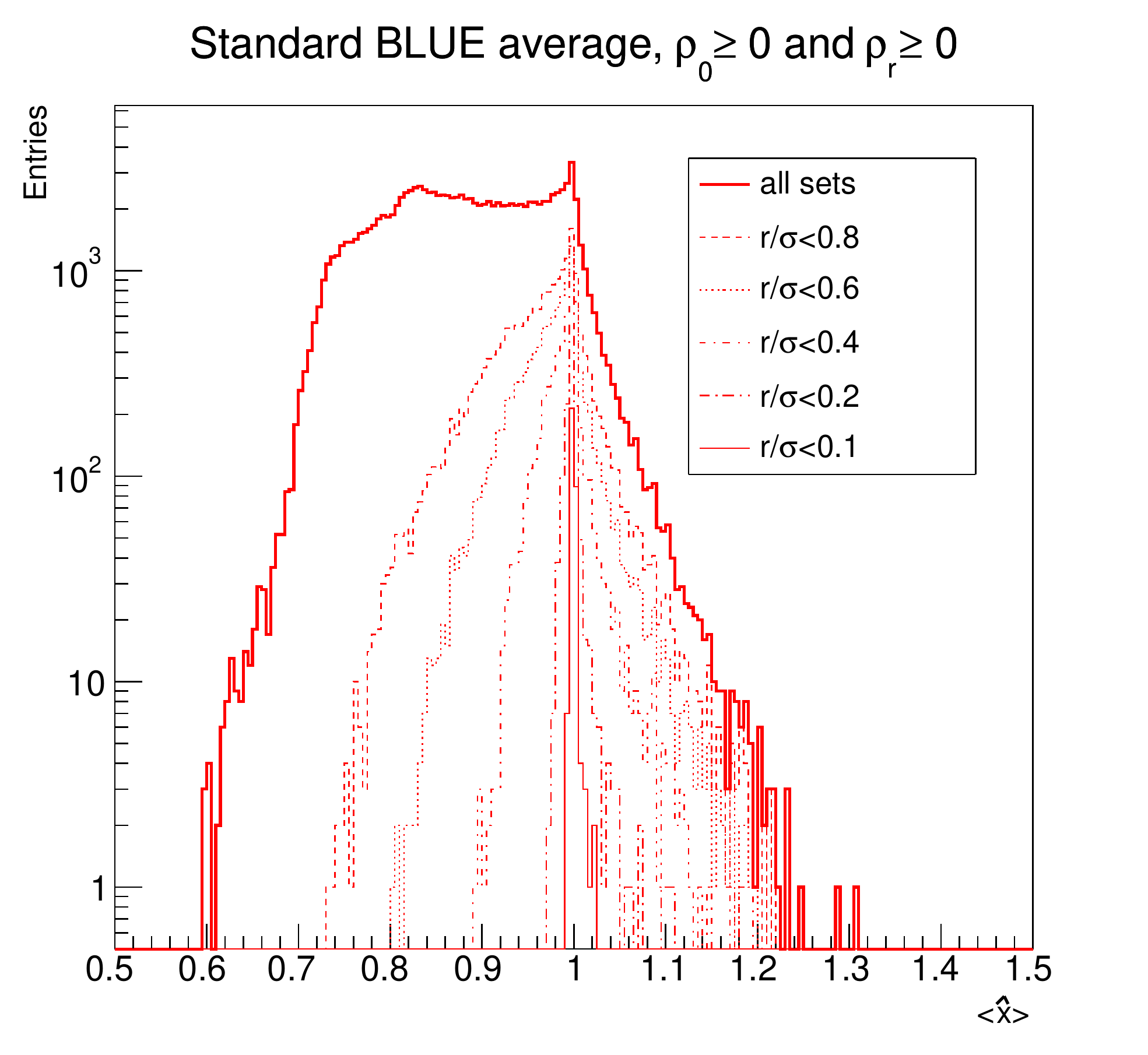}
 \includegraphics[width=0.49\textwidth]{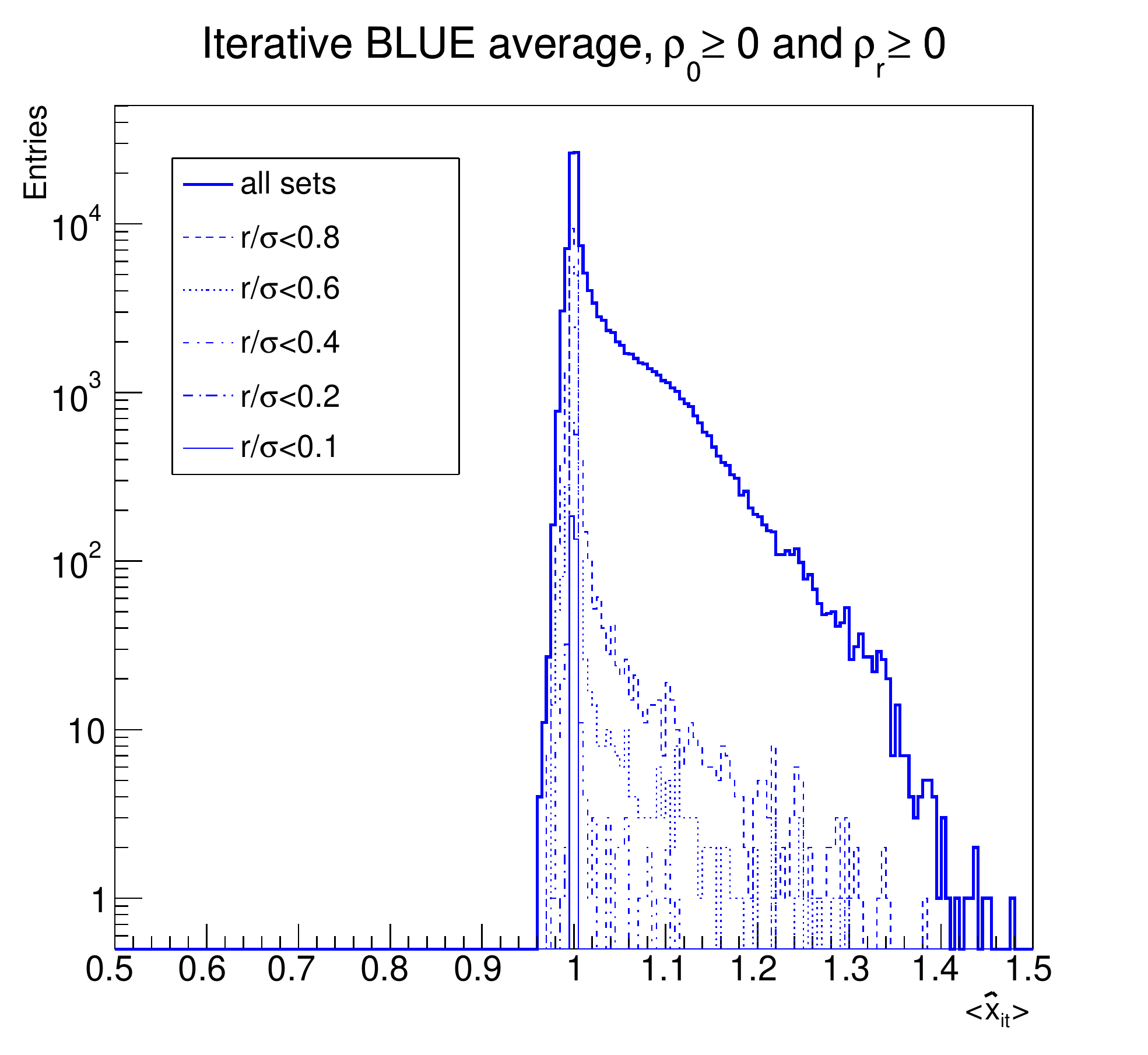}
  \includegraphics[width=0.49\textwidth]{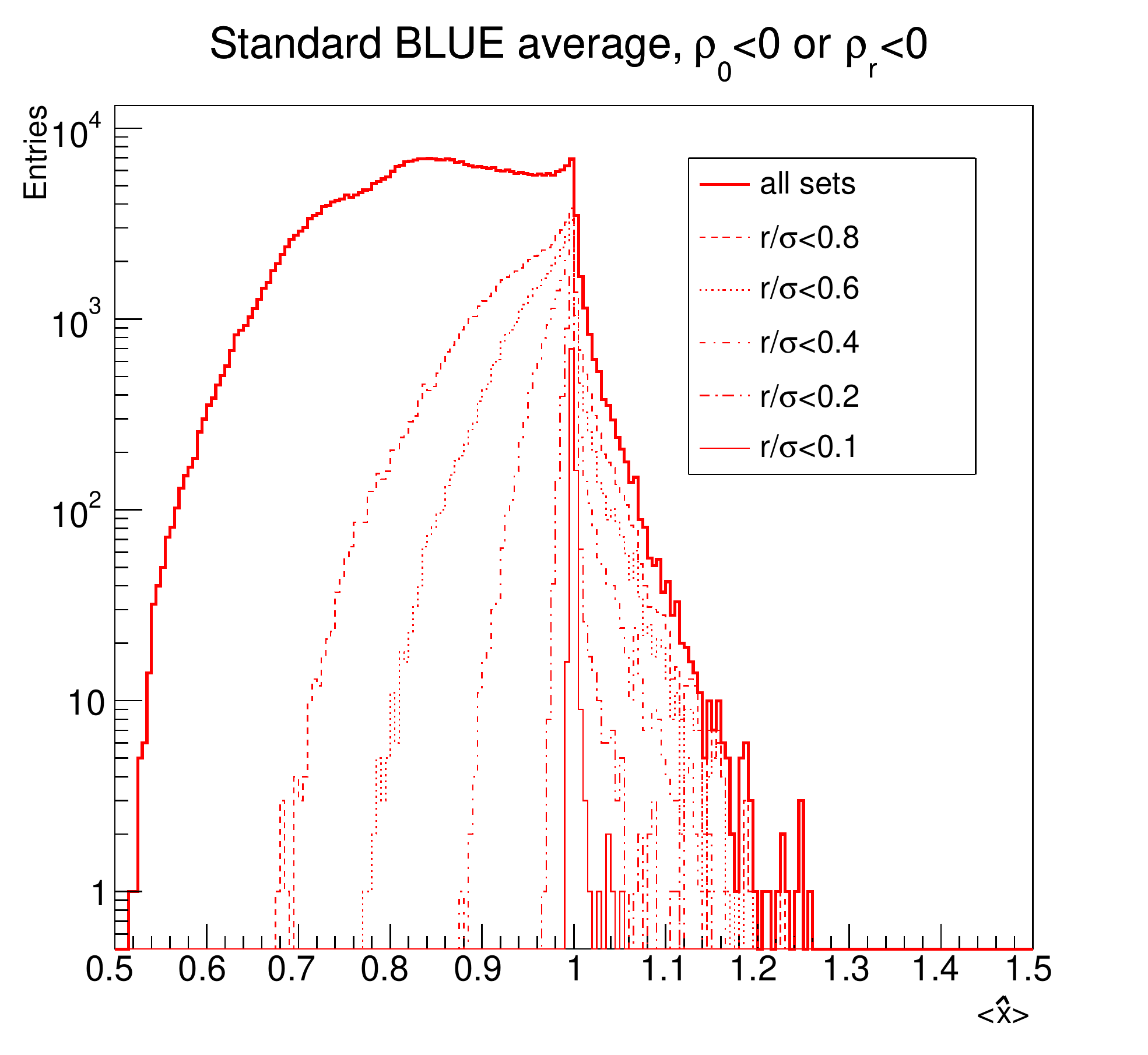}
 \includegraphics[width=0.49\textwidth]{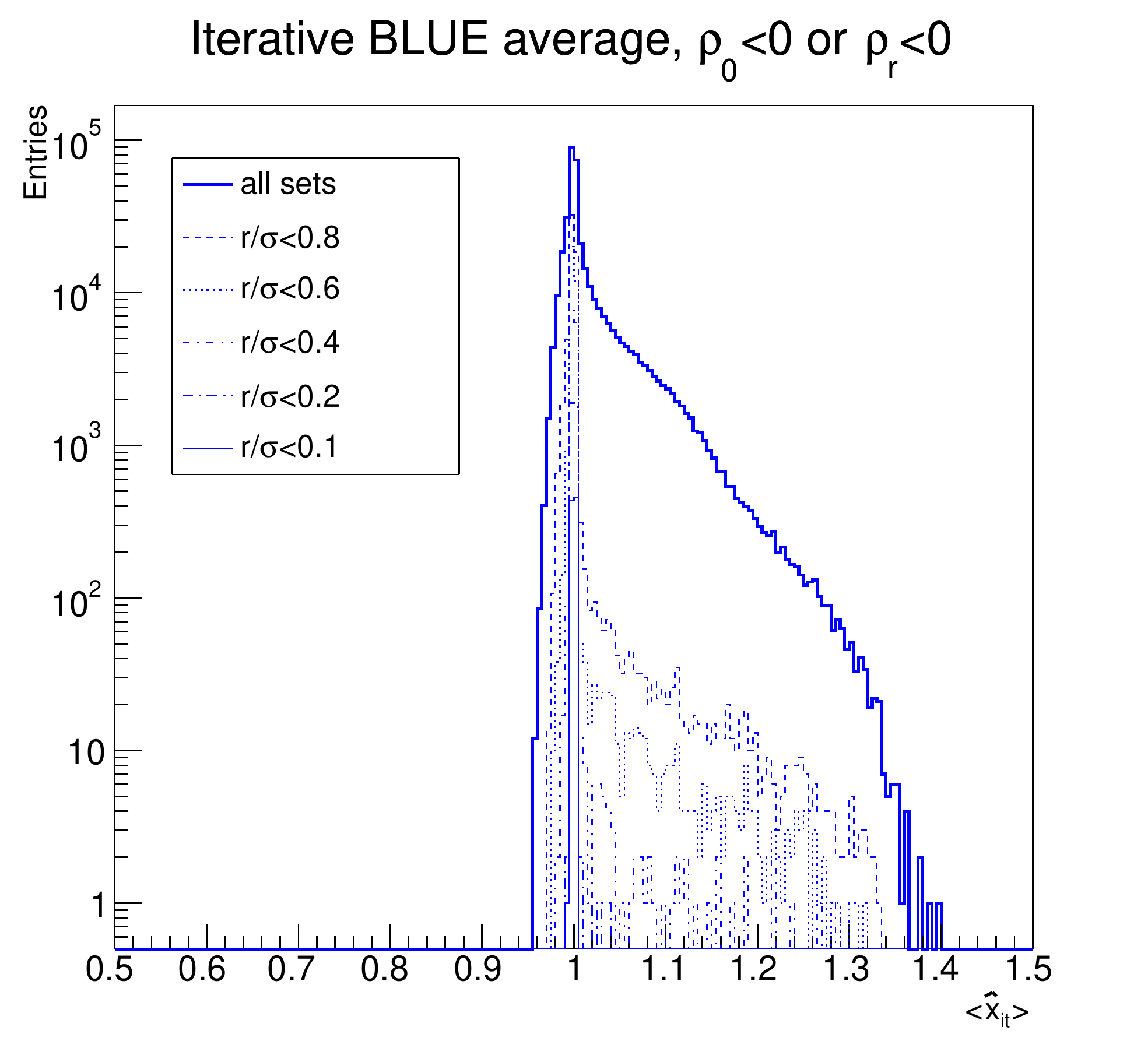}
 \caption{Distribution of the average value of the standard  (left) and iterative (right) BLUE estimates for different limits on $r/\sigma$
   for $\rho_0\ge 0$ and $\rho_{\mathrm{r}}\ge 0$ (top) and for $\rho_0< 0$ or $\rho_{\mathrm{r}}<0$ (bottom).}
\label{fig:bias-avg2}
\end{center}
\end{figure}

Figures~\ref{fig:bias-avg1} and~\ref{fig:bias-avg2} show the distributions of the average values $\left<\hat{x}\right>$ and $\left<\hat{x}_{\mathrm{it}}\right>$
of the standard and the iterative BLUE estimates, respectively, for all the simulated parameter sets and for sets with different 
upper bounds on  $r_1$ and $r_2$ or on $r_1/\sigma_1$ and $r_2/\sigma_2$. Plots are reported separately for 
the cases where both $\rho_0\ge 0$ and $\rho_{\mathrm{r}}\ge 0$ and where either $\rho_0<0$ or $\rho_{\mathrm{r}}<0$.
In these plots and in the following the variables subject to bounds are indicated for simplicity as $r$ and $r/\sigma$,
dropping the subscript 1 or 2. 
The shoulder with a local maximum around $\left<\hat{x}\right>= 0.8$ for the standard BLUE estimator is present because the 
uniform random sample of the parameter space is enriched in parameter sets where at least one uncertainty contribution is above 50\%,
which produce larger bias. This shoulder drops when upper bounds on $r$ or $r/\sigma$ are required.

The plots show that for $r/\sigma <0.2$ or $r<0.2$
the bias of the standard method ranges from $-4\%$ up to, for a very small number of cases, $+10\%$,
while the bias of the iterative method remains below one or few percent in most of the case.
In general, the bias of the iterative method is significantly smaller than the standard method for a large majority of the cases,
though it may still exhibits large values in a limited fraction of the cases.

\begin{figure}[htpb]
 \begin{center}
  \includegraphics[width=0.49\textwidth]{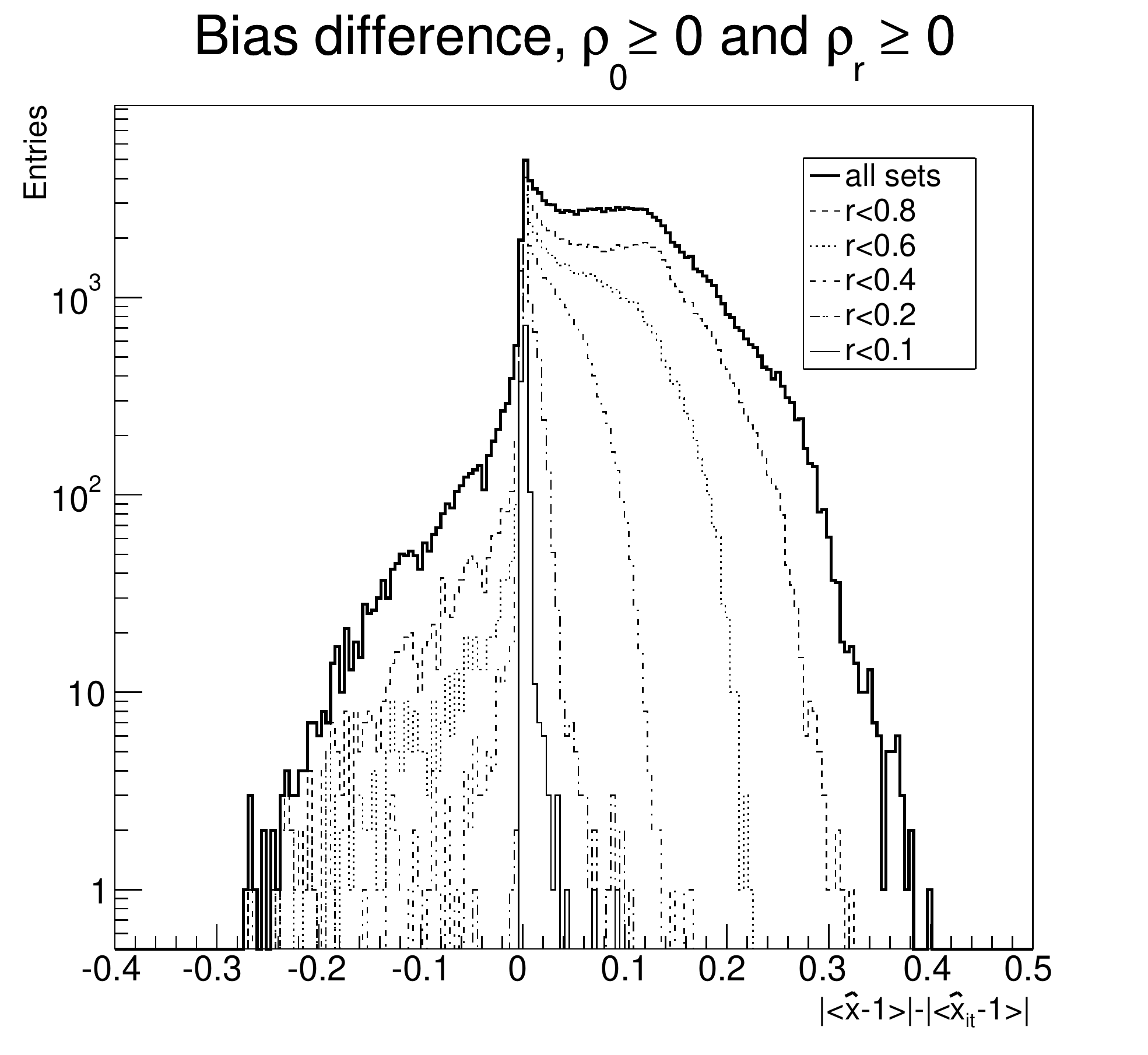}
 \includegraphics[width=0.49\textwidth]{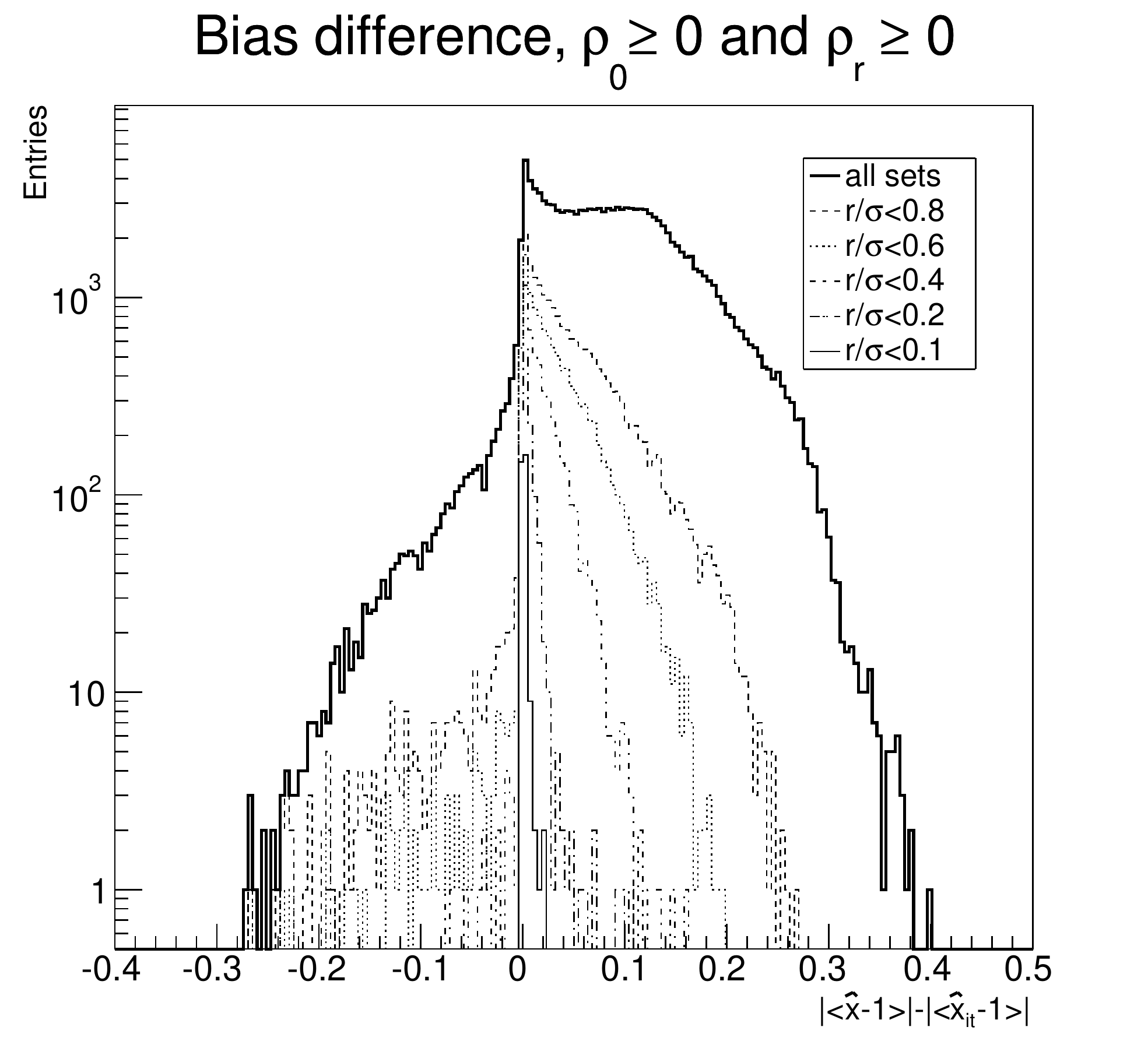}
  \includegraphics[width=0.49\textwidth]{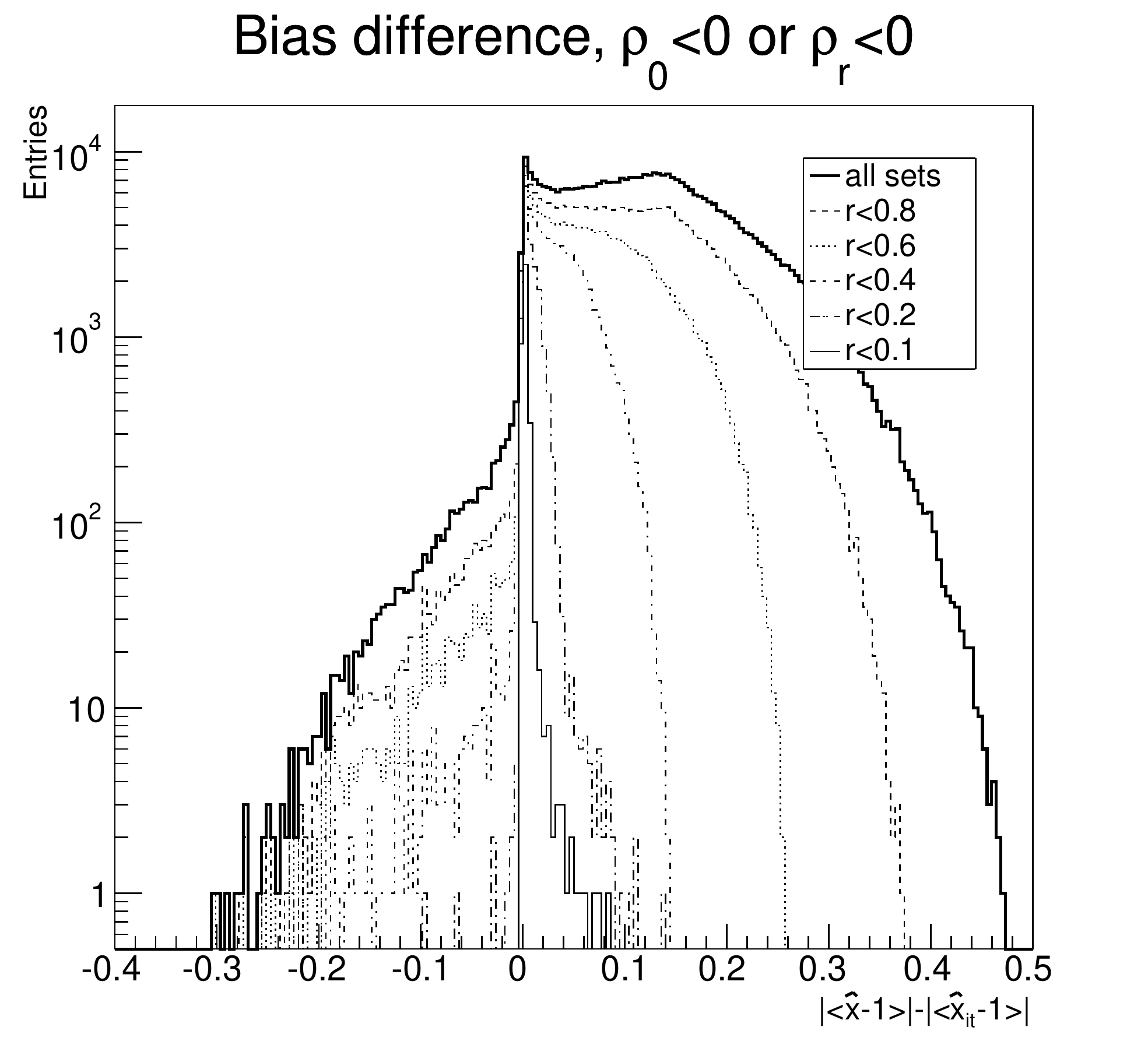}
 \includegraphics[width=0.49\textwidth]{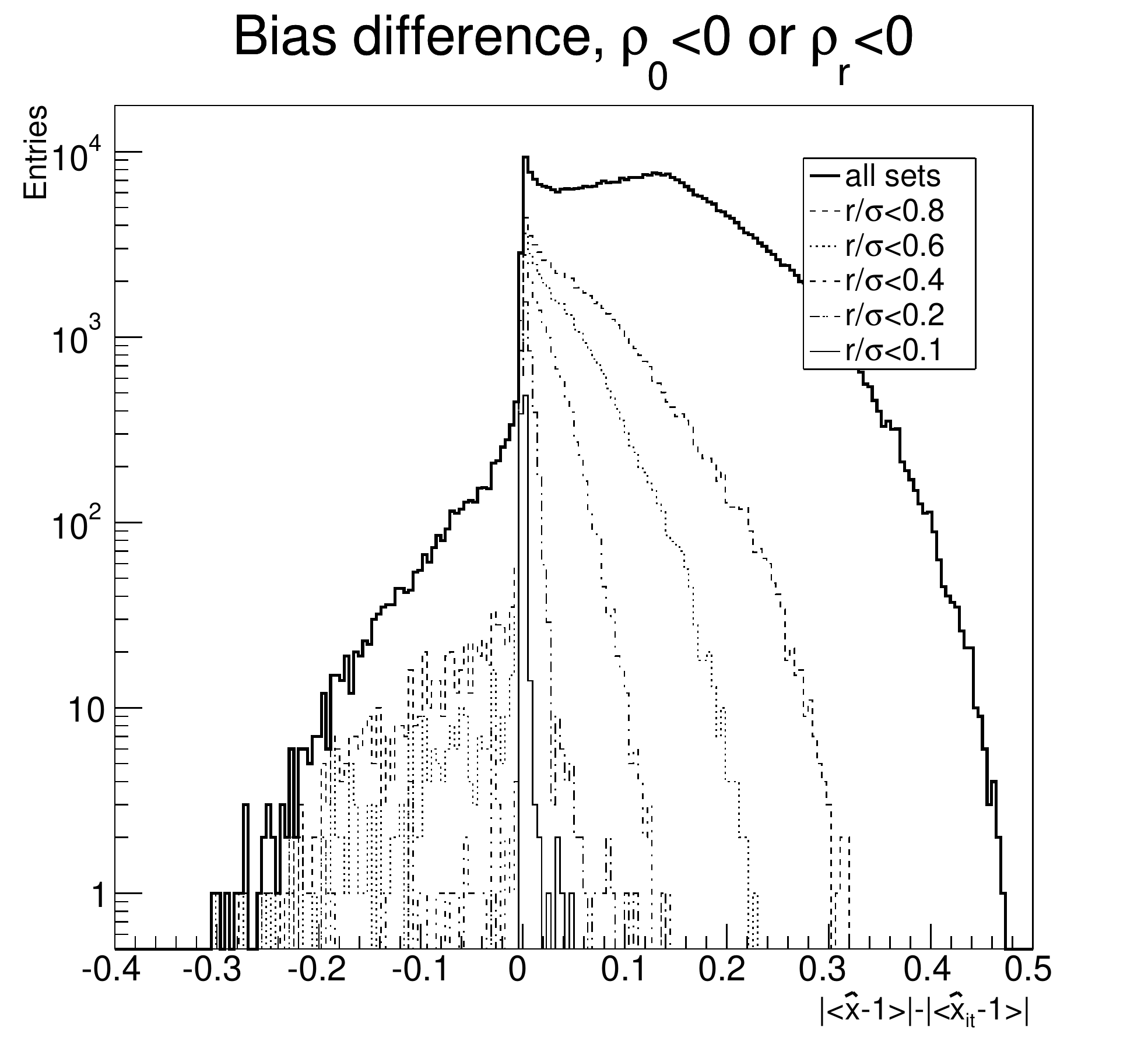}
 \caption{Difference of measured absolute value of the bias for the standard and iterative BLUE estimates for different limits on $r$
(left) and for different limits on $r/\sigma$ (right)  for $\rho_0\ge 0$ and $\rho_{\mathrm{r}}\ge 0$ (top) and for $\rho_0< 0$ or $\rho_{\mathrm{r}} <0$ (bottom).
Positive values indicate a smaller bias in the iterative method compared to the standard method.}
\label{fig:bias-avgd-vsr}
\end{center}
\end{figure}
Figure~\ref{fig:bias-avgd-vsr} shows the distribution of the difference of
the absolute values of the biases for standard and the iterative BLUE estimates,  
$\left<\hat{x}-x_0\right>$ and $\left<\hat{x}_{\mathrm{it}}-x_0\right>$,
for all the simulated parameter sets and for the sets with different upper bounds on $r$ or $r/\sigma$,
again separately for $\rho_0\ge 0$ and $\rho_{\mathrm{r}}\ge 0$ and for either $\rho_0<0$ or $\rho_{\mathrm{r}}<0$.
The bias of the two methods is identical within one percent
for $r<0.1$ or $r/\sigma<0.1$ for most of the cases, and the difference remains less than 4\% for most of the cases
with $r<0.2$ or $r/\sigma<0.2$.
The iterative method appears to have a smaller or almost identical bias compared to
the standard method in the vast majority of the cases.
There are cases where the bias of the standard methods is smaller than the bias of the iterative method, but this 
tends to happens only when either $r$ or $r/\sigma$ is very large.

\begin{figure}[htpb]
 \begin{center}
 \includegraphics[width=0.49\textwidth]{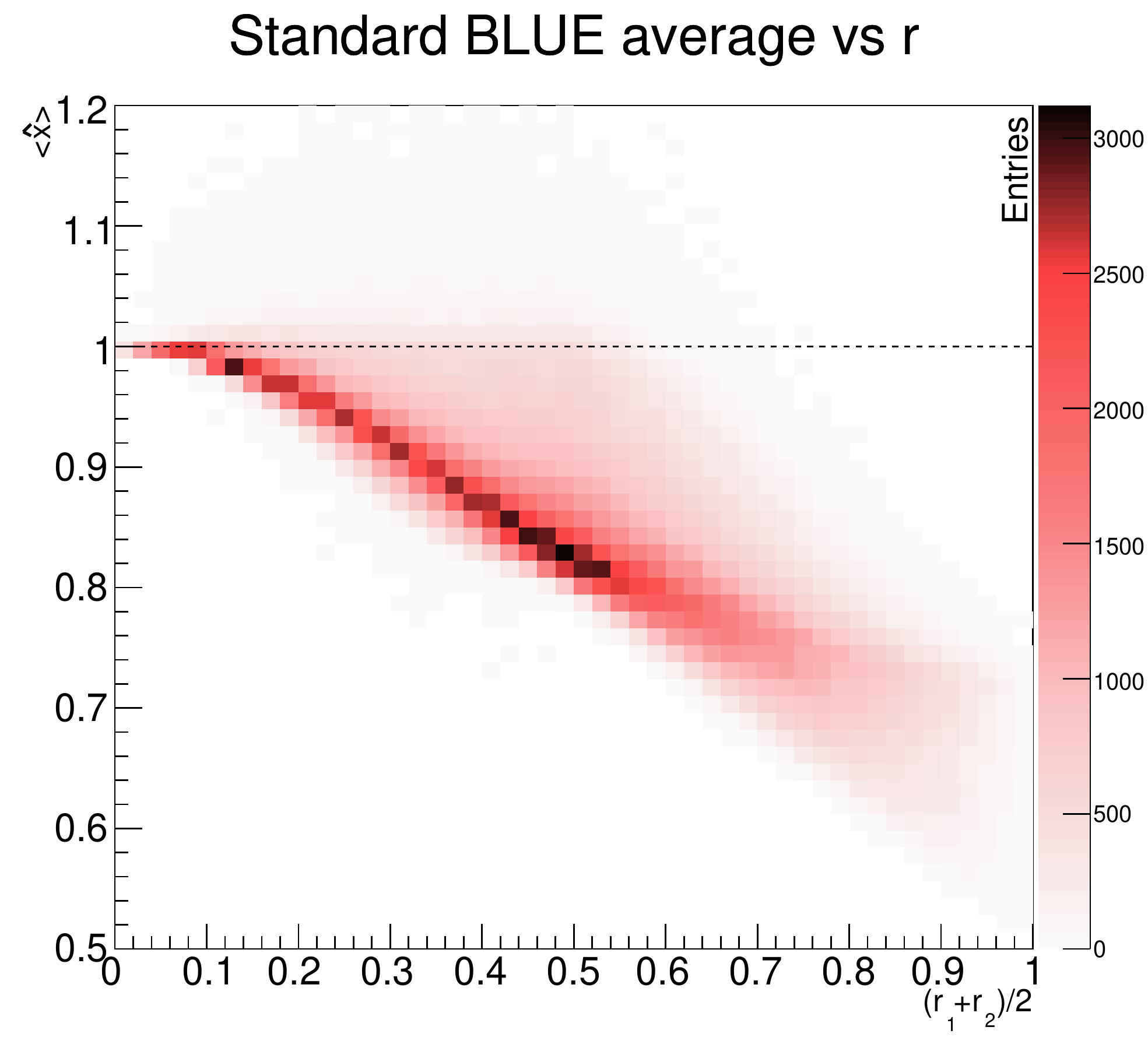}
 \includegraphics[width=0.49\textwidth]{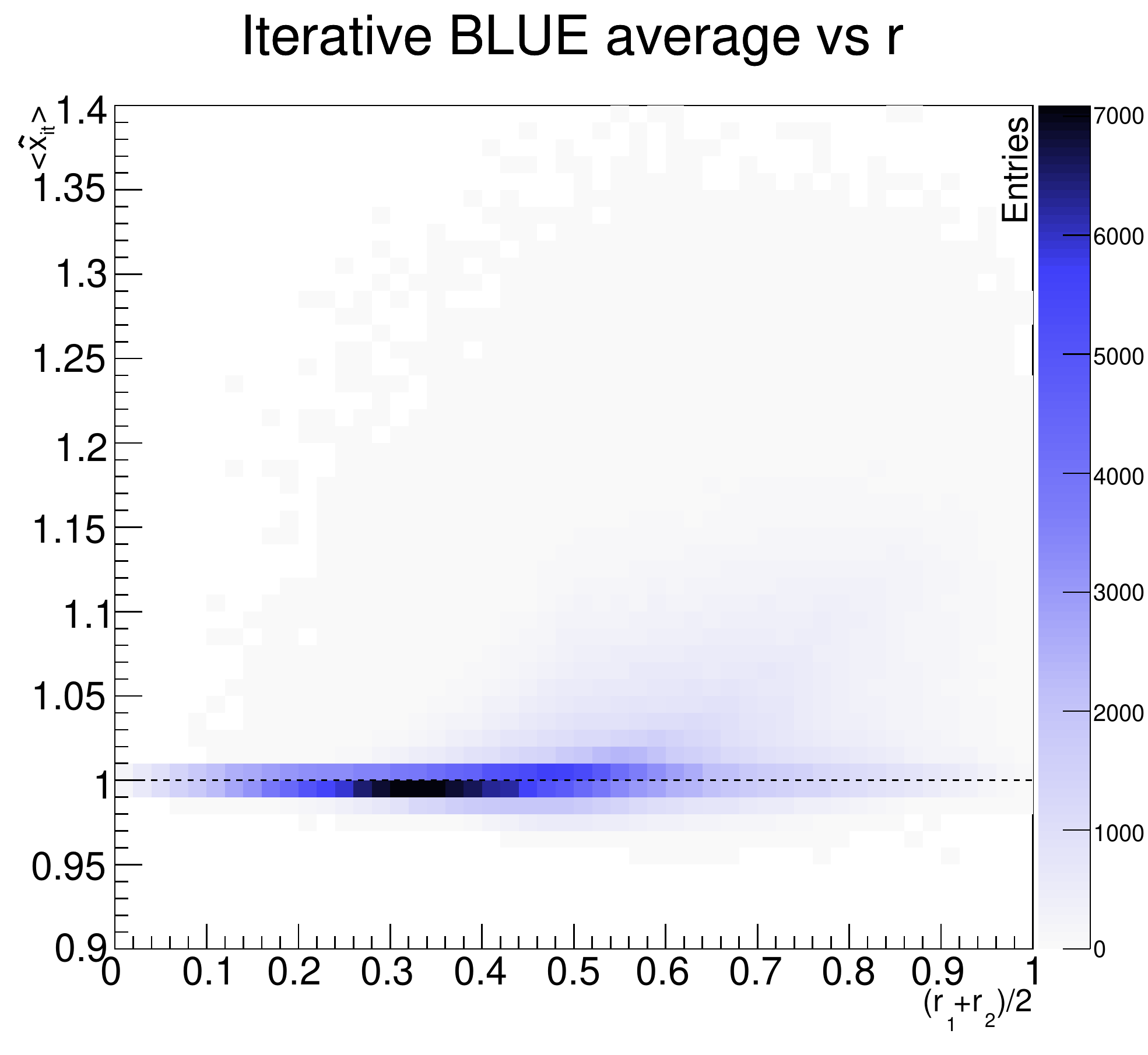}
 \caption{Distribution of the average value measured with the standard (left) and iterative (right) BLUE method as a function of $\frac{r_1+r_2}{2}$.}
\label{fig:bias-avg-rr}
\end{center}
\end{figure}

\begin{figure}[htpb]
 \begin{center}
 \includegraphics[width=0.49\textwidth]{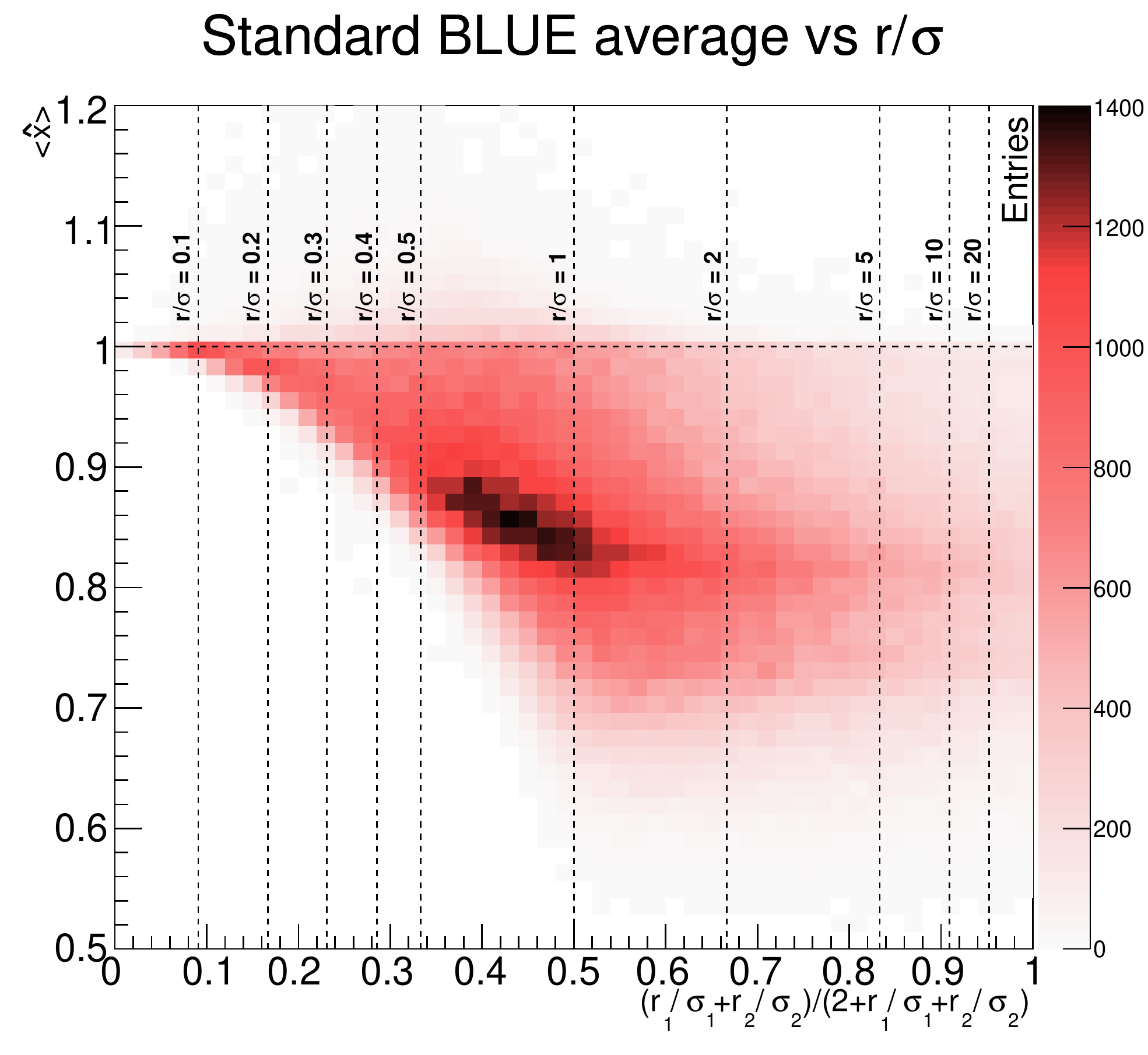}
 \includegraphics[width=0.49\textwidth]{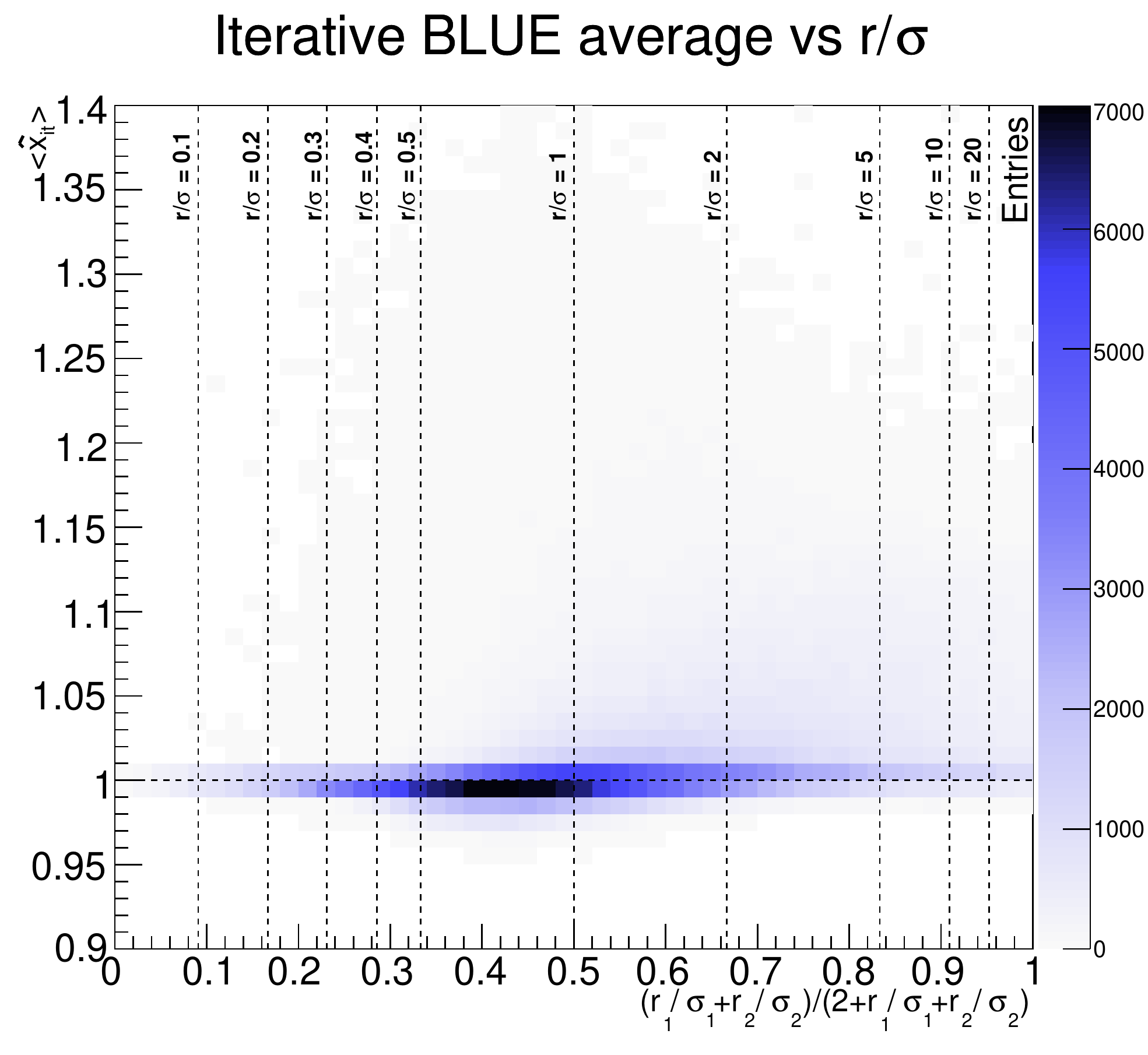}
 \caption{Distribution of the average value measured with the standard (left) and iterative (right) BLUE method as a function
of $\frac{r_1/\sigma_1+r_2/\sigma_2}{2+r_1/\sigma_2+r_2/\sigma_2}$, which is a rescaling of $\frac{r_1/\sigma_1+r_2/\sigma_2}{2}$.}
\label{fig:bias-avg-r}
\end{center}
\end{figure}

\begin{figure}[htpb]
 \begin{center}
 \includegraphics[width=0.49\textwidth]{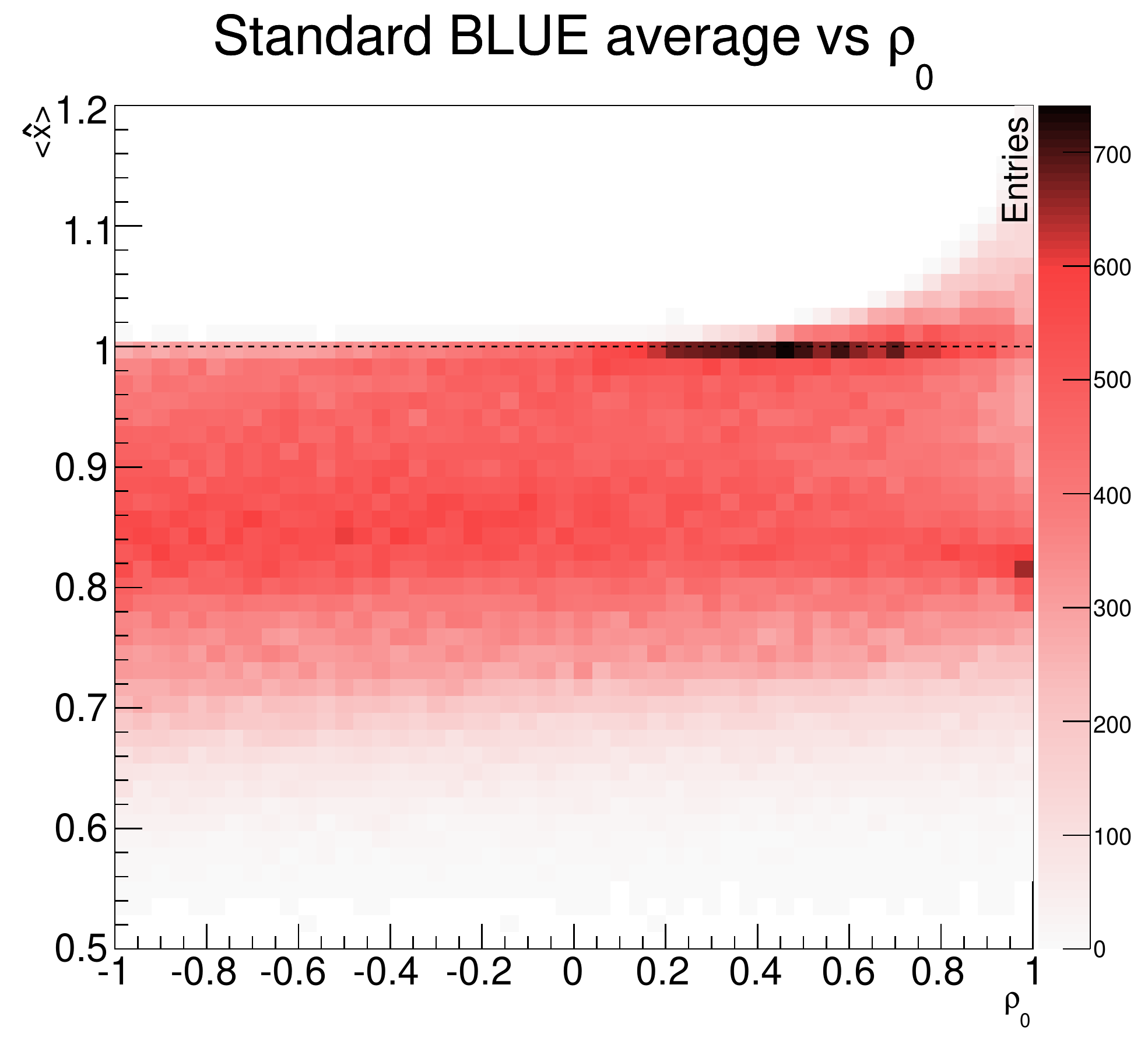}
 \includegraphics[width=0.49\textwidth]{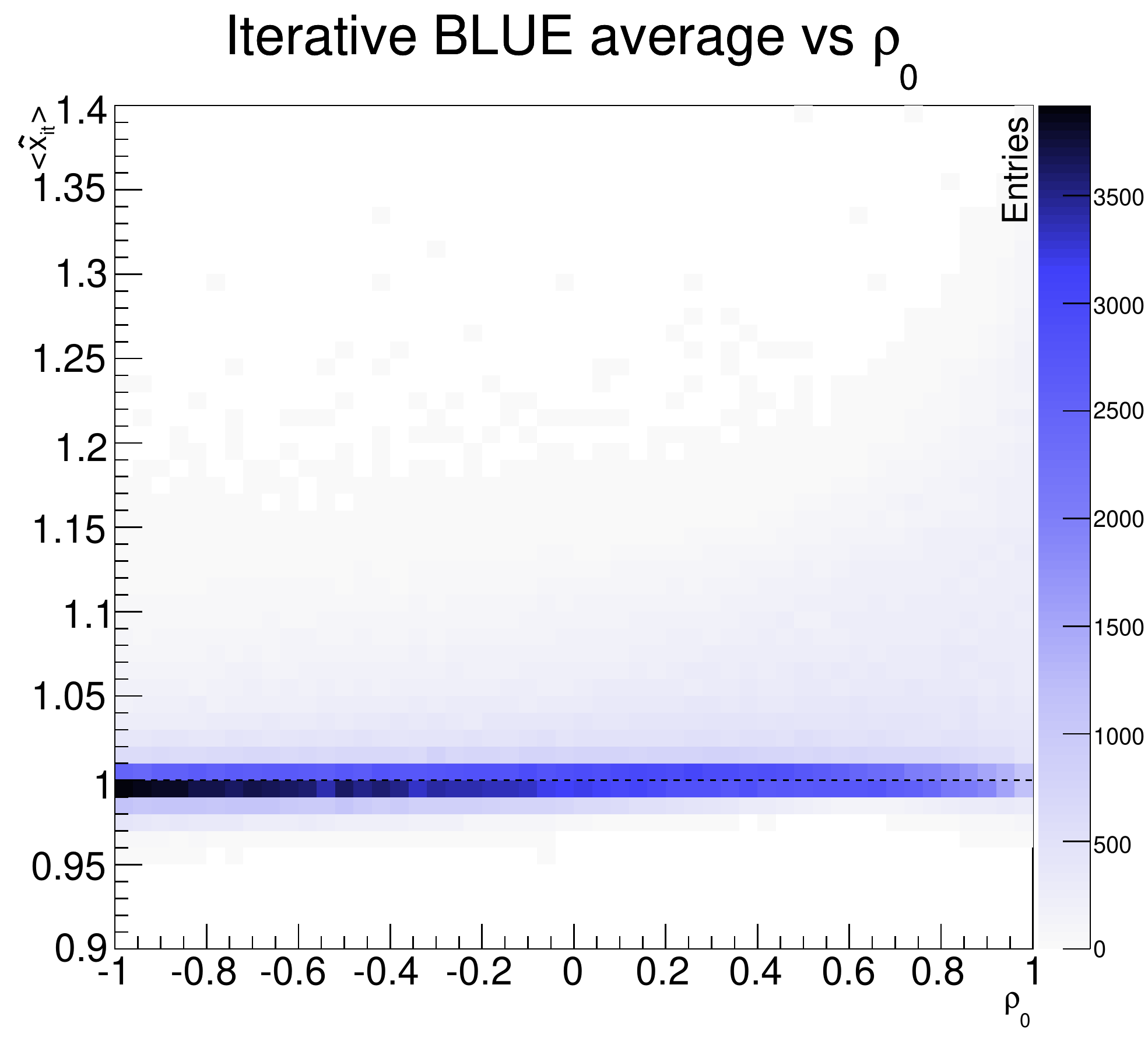}
 \caption{Distribution of the average value measured with the standard (left) and iterative (right) BLUE method as a function of $\rho_0$.}
\label{fig:bias-avg-rho}
\end{center}
\end{figure}

\begin{figure}[htpb]
 \begin{center}
 \includegraphics[width=0.49\textwidth]{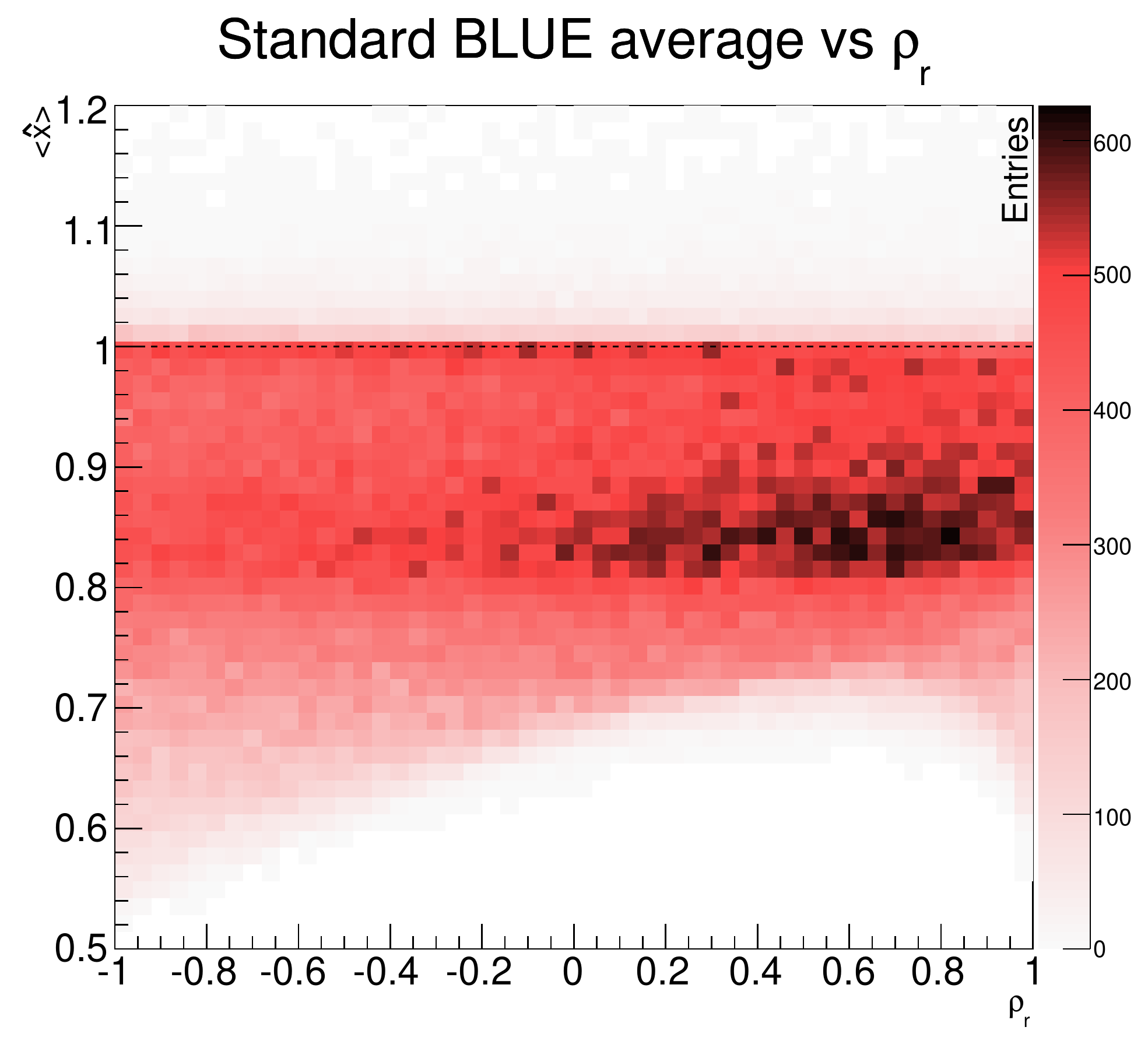}
 \includegraphics[width=0.49\textwidth]{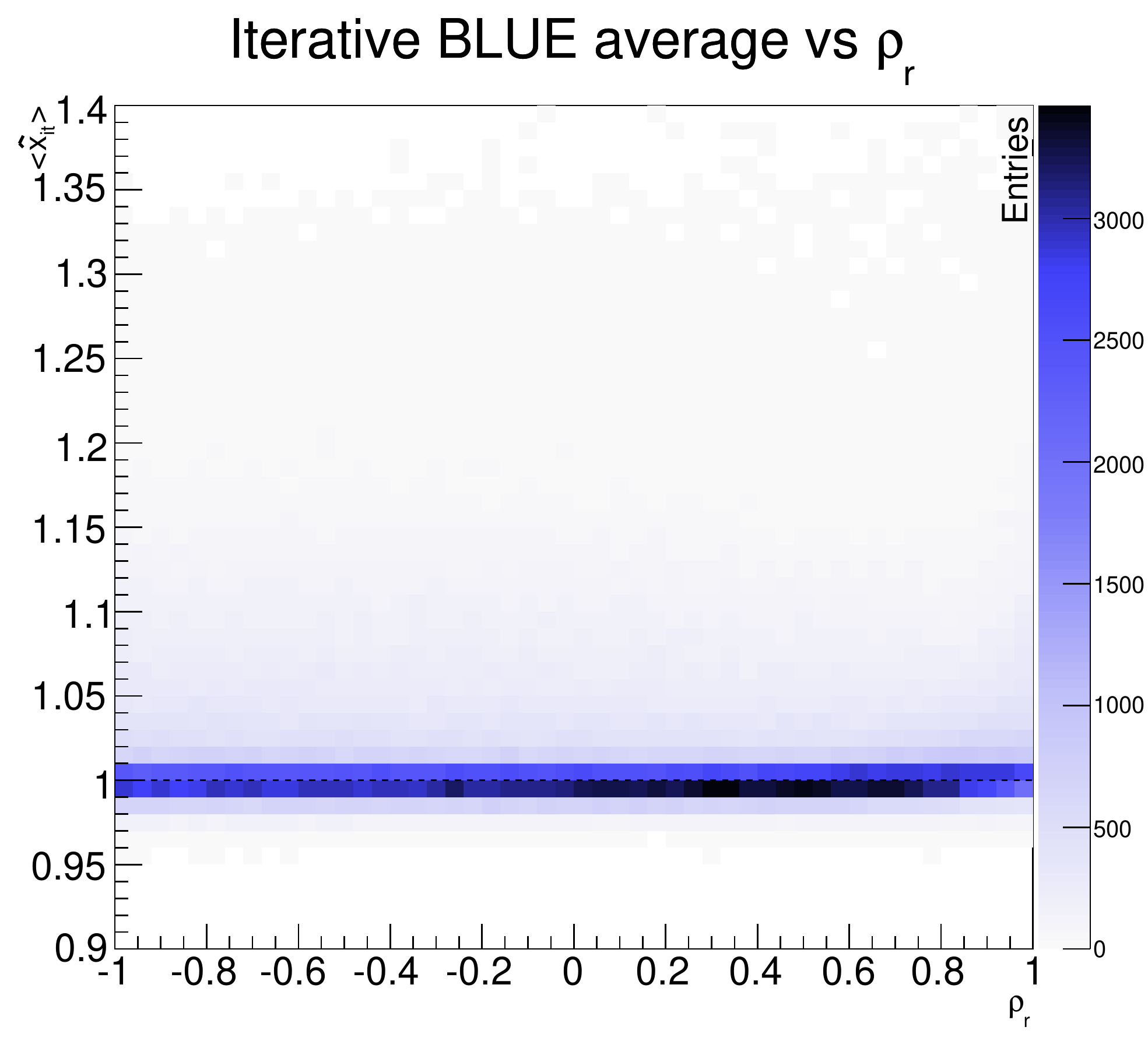}
 \caption{Distribution of the average value measured with the standard (left) and iterative (right) BLUE method as a function of $\rho_{\mathrm{r}}$.}
\label{fig:bias-avg-rhor}
\end{center}
\end{figure}

\begin{figure}[htpb]
 \begin{center}
 \includegraphics[width=0.49\textwidth]{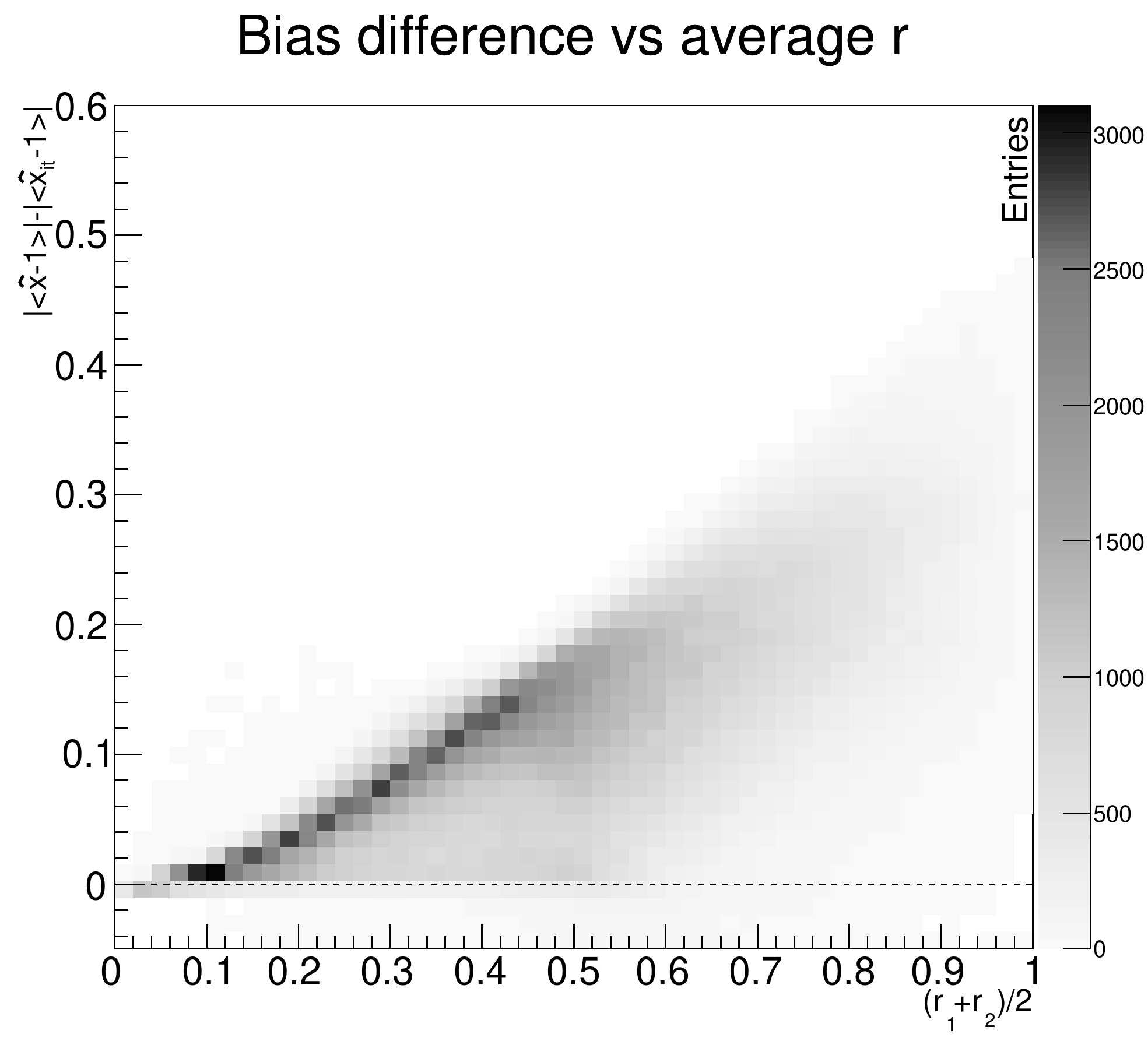}
 \includegraphics[width=0.49\textwidth]{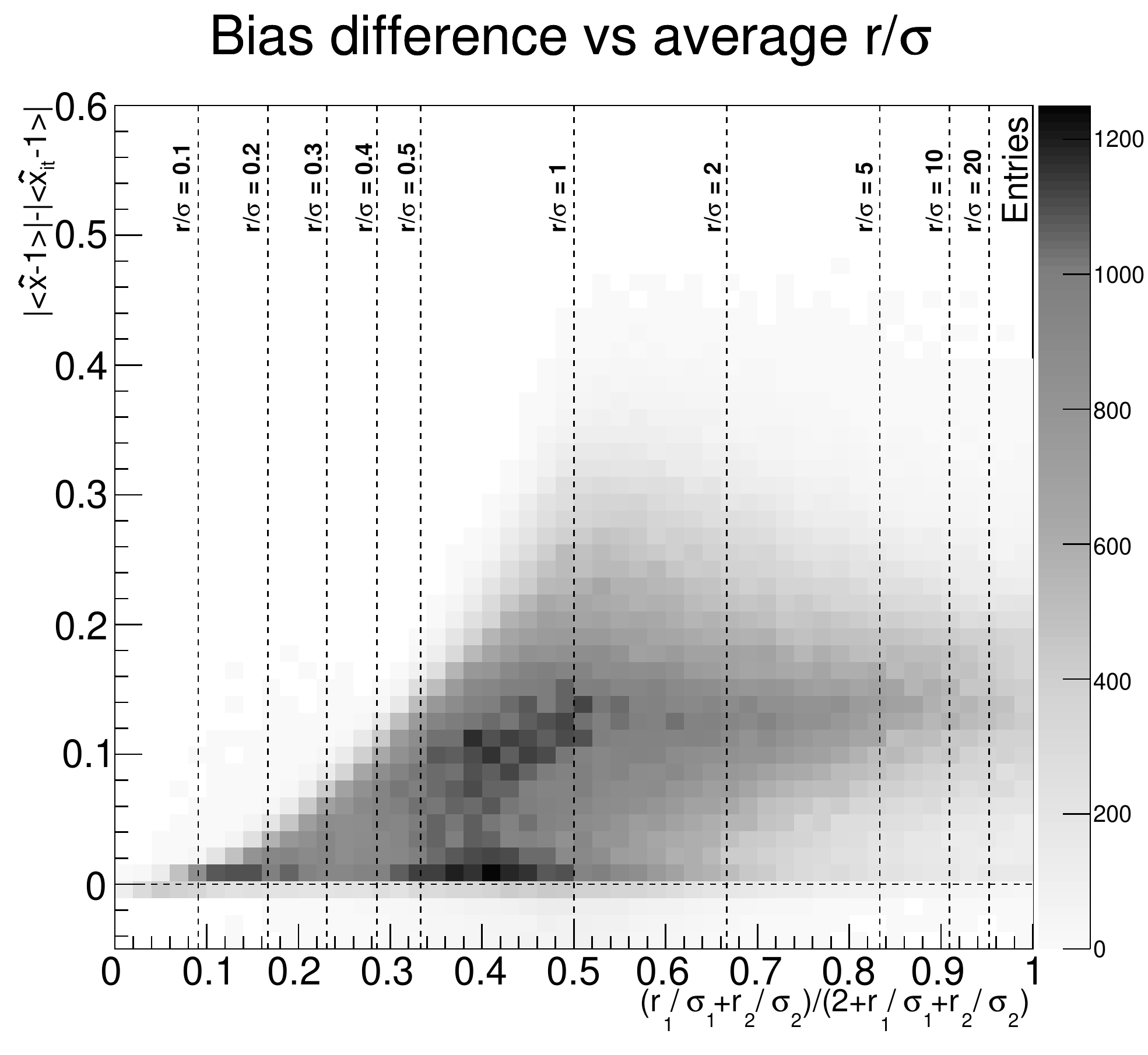}
 \includegraphics[width=0.49\textwidth]{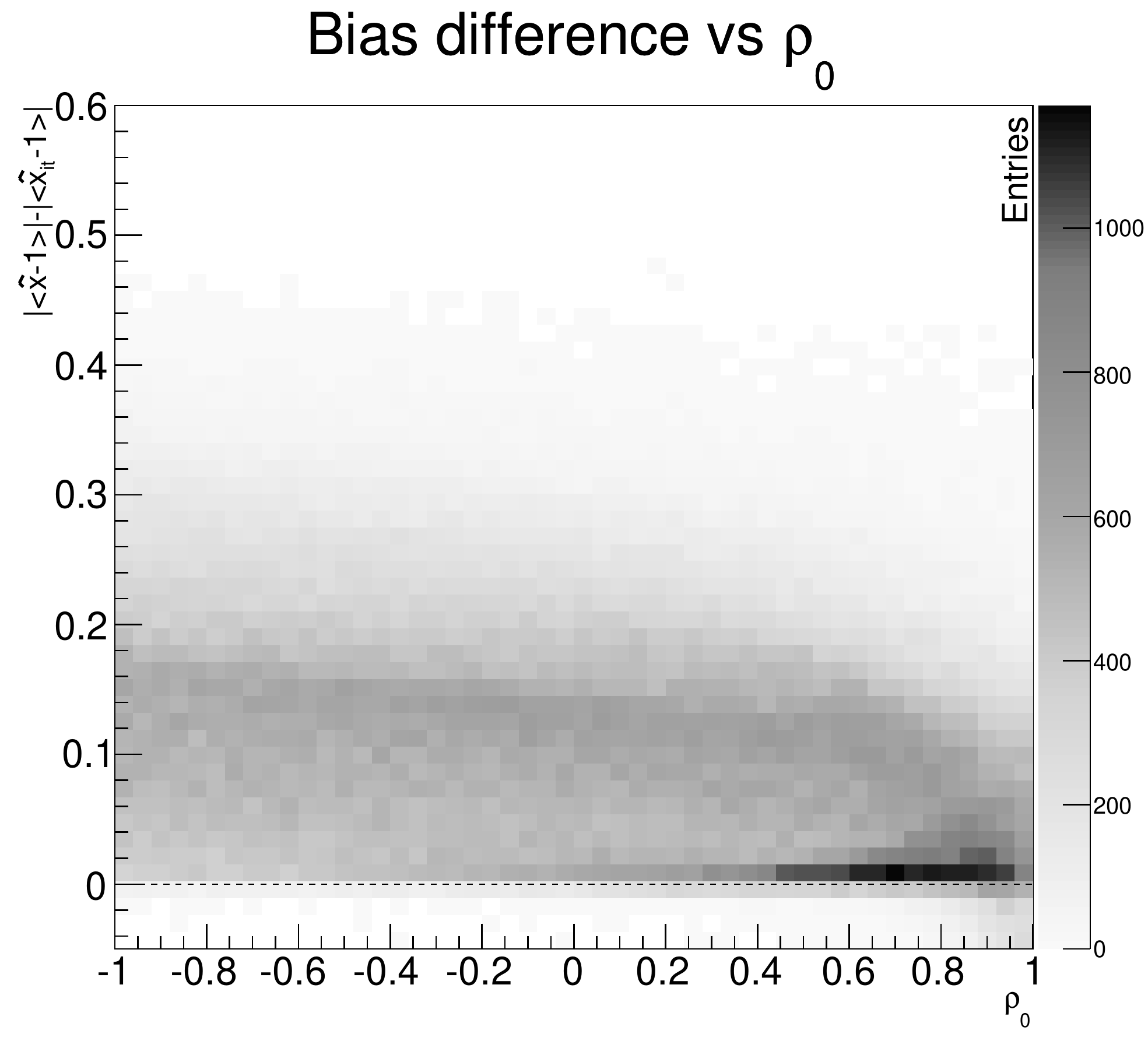}
 \includegraphics[width=0.49\textwidth]{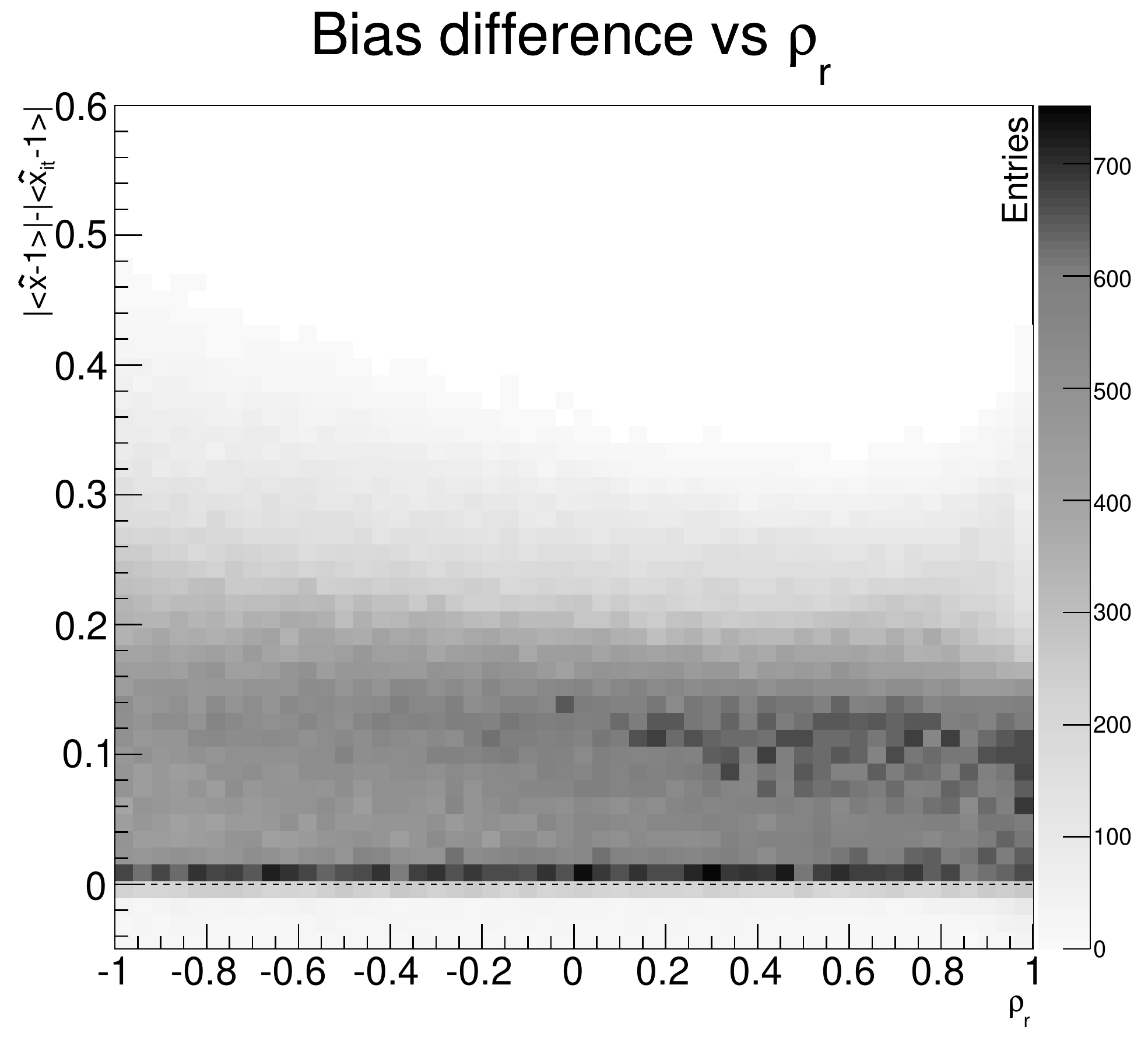}
 \caption{Difference of measured absolute value of the bias for the standard and iterative BLUE methods 
as a function of $\frac{r_1+r_2}{2}$ (top left) and $\frac{r_1/\sigma_1+r_2/\sigma_2}{2+r_1/\sigma_2+r_2/\sigma_2}$ (top right),
$\rho_0$ (bottom left) and $\rho_{\mathrm{r}}$ (bottom right).}
\label{fig:bias-avgd}
\end{center}
\end{figure}

Figures~\ref{fig:bias-avg-rr}, \ref{fig:bias-avg-r}, \ref{fig:bias-avg-rho} and \ref{fig:bias-avg-rhor}
show the distributions of the average values $\left<\hat{x}\right>$ and $\left<\hat{x}_{\mathrm{it}}\right>$ of the standard and the iterative BLUE estimators, respectively, as a function of the
average of the two relative contributions $r_1$ and $r_2$: $\frac{r_1+r_2}{2}$, as a function of
$\frac{r_1/\sigma_1+r_2/\sigma_2}{2+r_1/\sigma_2+r_2/\sigma_2}$, which is a convenient way to rescale
$\frac{r_1/\sigma_1+r_2/\sigma_2}{2}$ in the interval $[0, 1]$,
and as a function of $\rho_0$ and $\rho_{\mathrm{r}}$.
Note that the vertical scales are different in the plots corresponding
to the standard and iterative methods.
Both  $r_i$ and $r_i/\sigma_i$, $i =1, 2$,
have large impact on the bias of the standard method, while $\rho_0$ and $\rho_{\mathrm{r}}$ have 
smaller impact on the averages, except for cases with very large correlation values.
The iterative BLUE estimator has always a much smaller sensitivity on all parameters in the present study
compared to the standard estimator.
Fig.~\ref{fig:bias-avgd} shows the distribution of the difference of the absolute values of the biases $\left<\hat{x}-x_0\right>$ and $\left<\hat{x}_{\mathrm{it}}-x_0\right>$ for standard and the iterative BLUE methods, respectively,
as a function of $\frac{r_1+r_2}{2}$, $\frac{r_1/\sigma_1+r_2/\sigma_2}{2+r_1/\sigma_2+r_2/\sigma_2}$, $\rho_0$ and $\rho_{\mathrm{r}}$.


\begin{figure}[htpb]
 \begin{center}
 \includegraphics[width=0.49\textwidth]{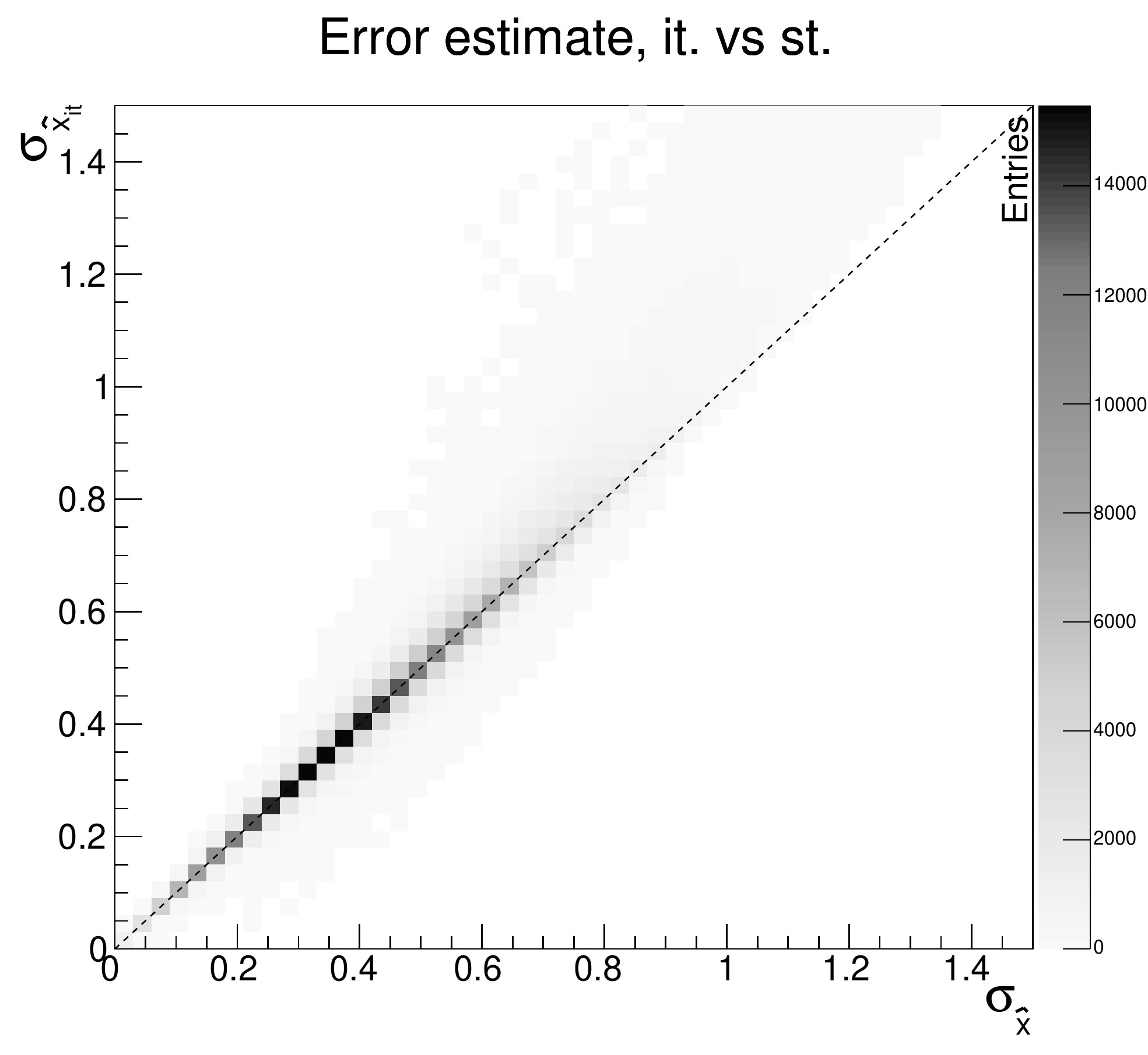}
  \caption{Distribution of the error estimates for the iterative versus standard the  BLUE methods for all generated sets.}
 \label{fig:rms}
\end{center}
\end{figure}
\begin{figure}[htpb]
 \begin{center}
 \includegraphics[width=0.49\textwidth]{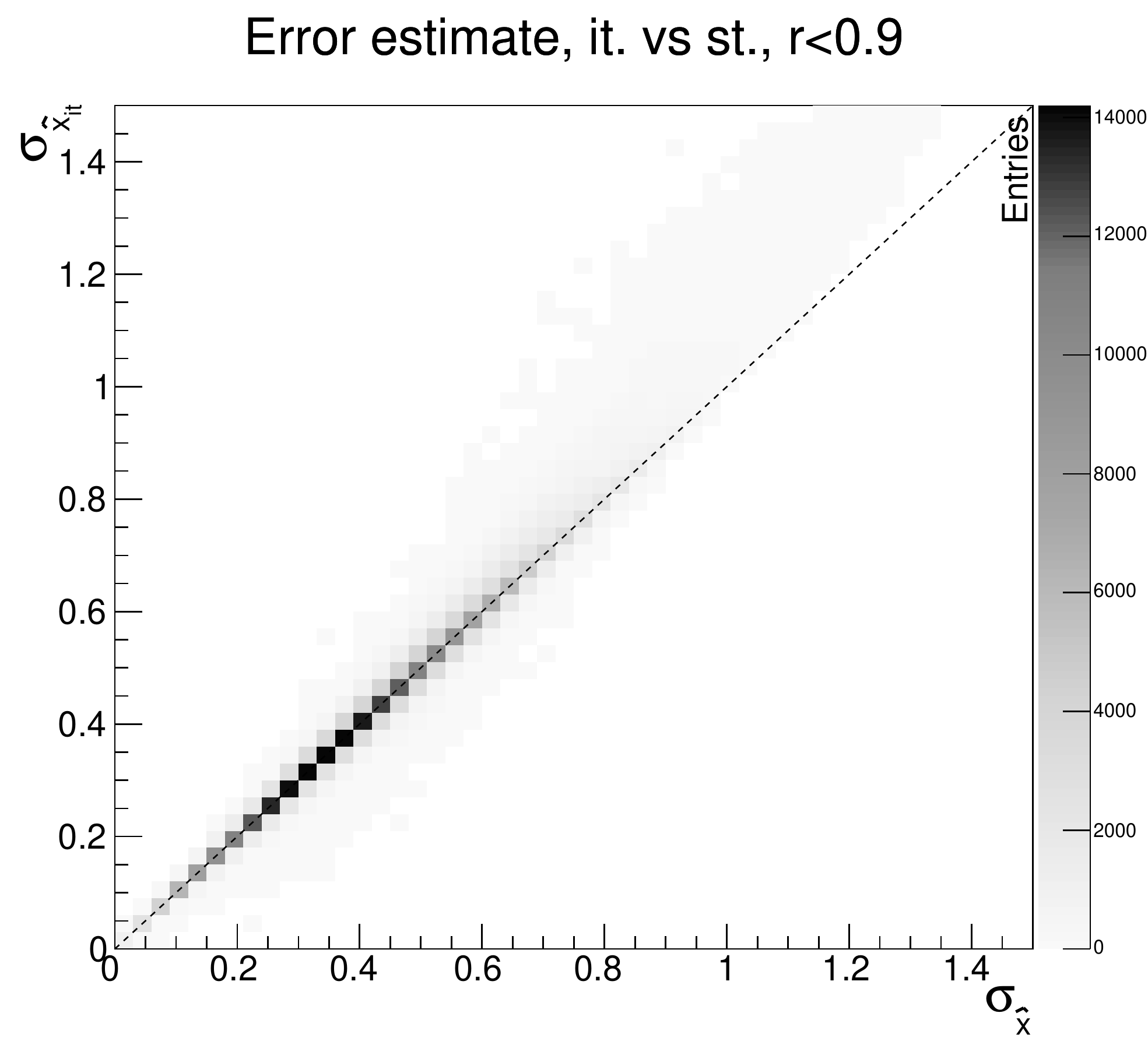}
 \includegraphics[width=0.49\textwidth]{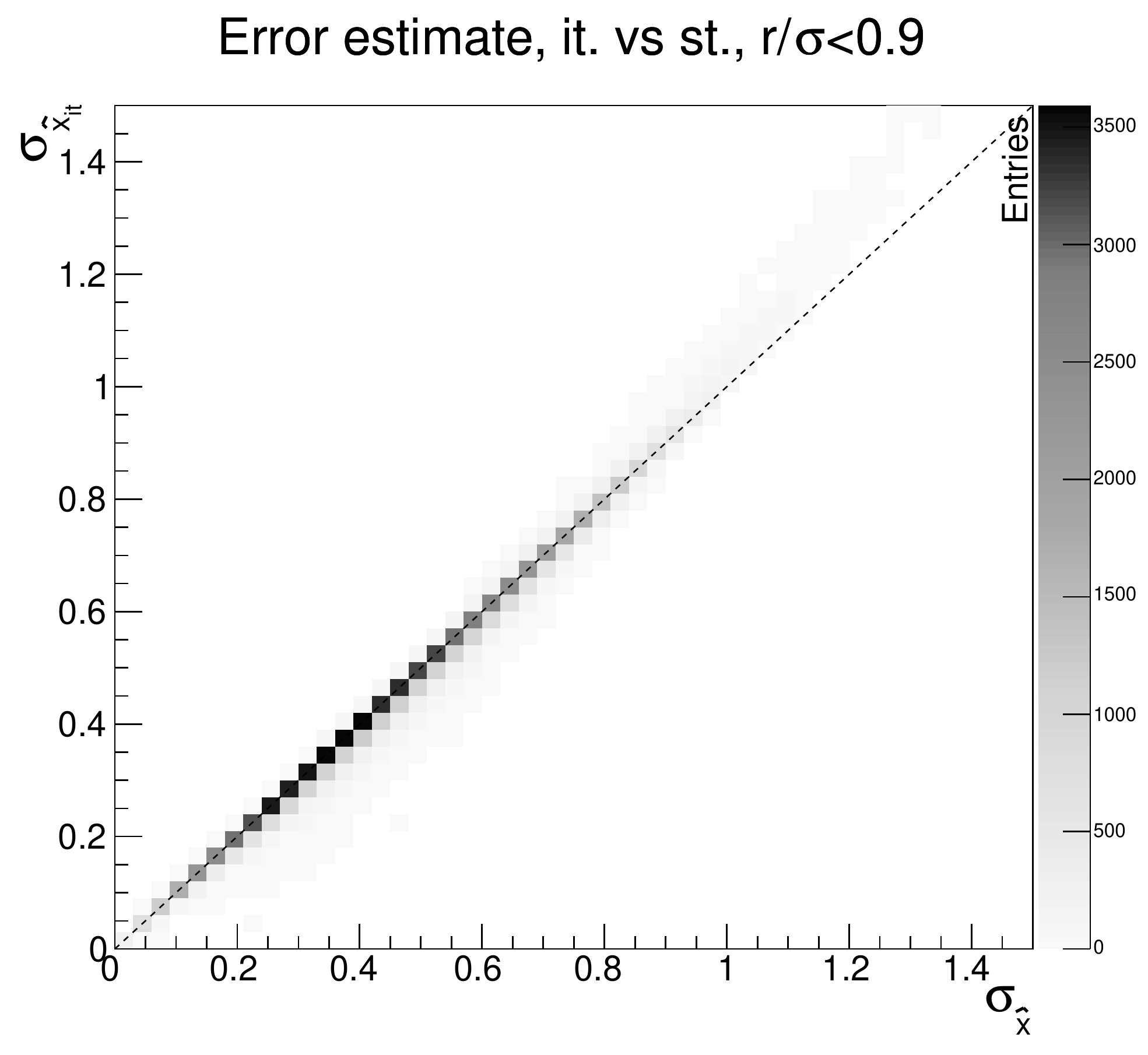}
 \includegraphics[width=0.49\textwidth]{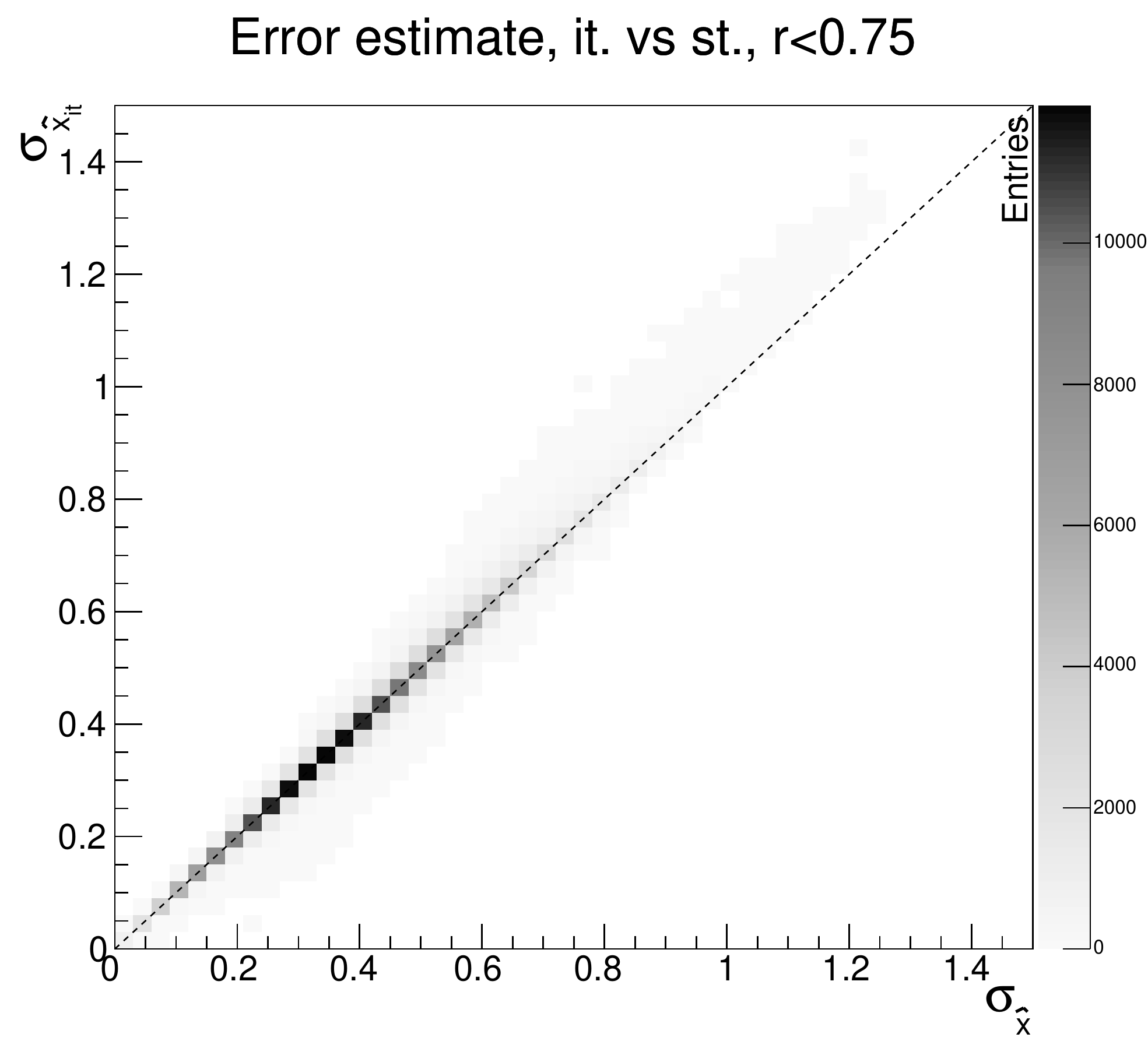}
  \includegraphics[width=0.49\textwidth]{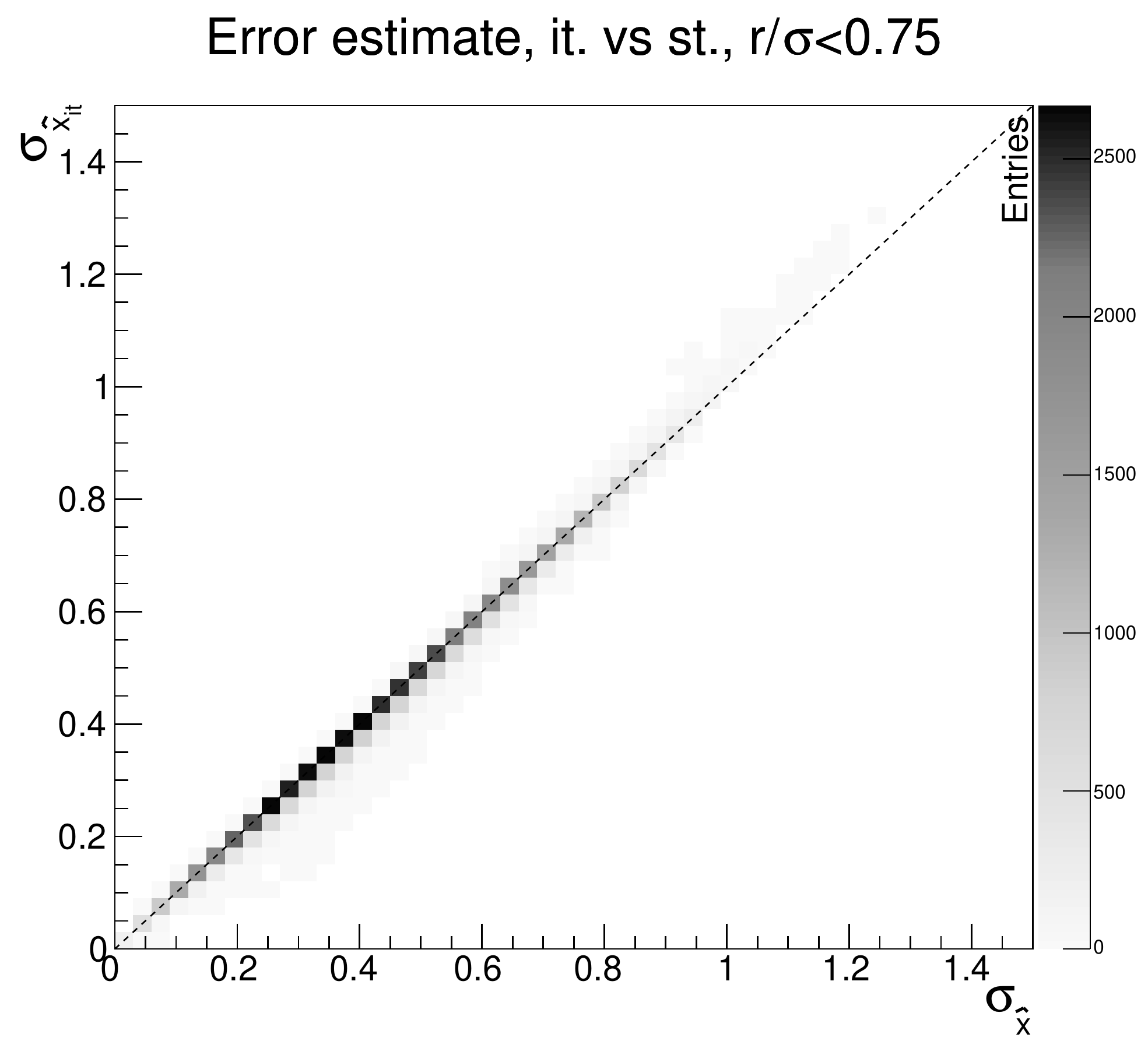}
 \includegraphics[width=0.49\textwidth]{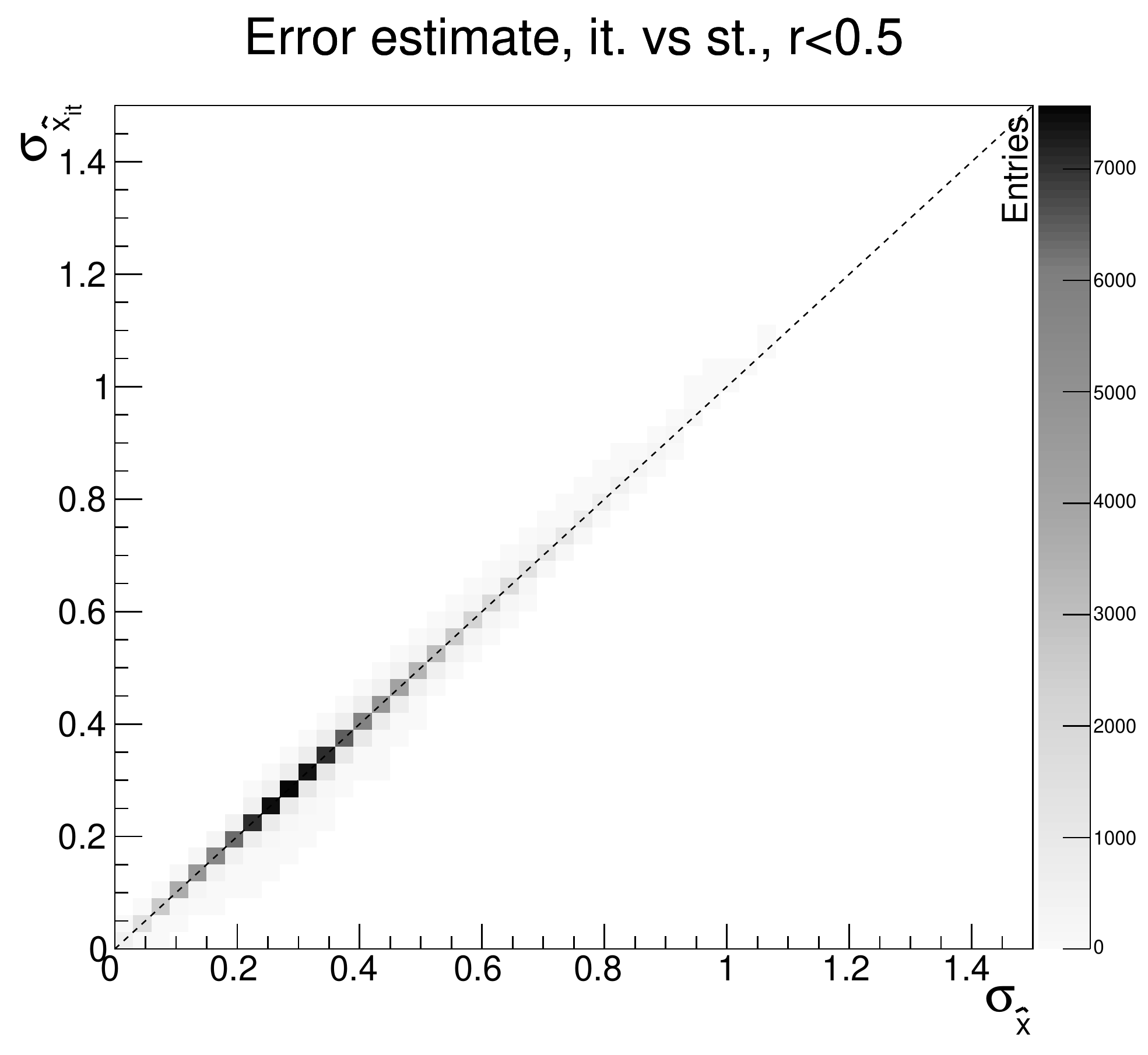}
 \includegraphics[width=0.49\textwidth]{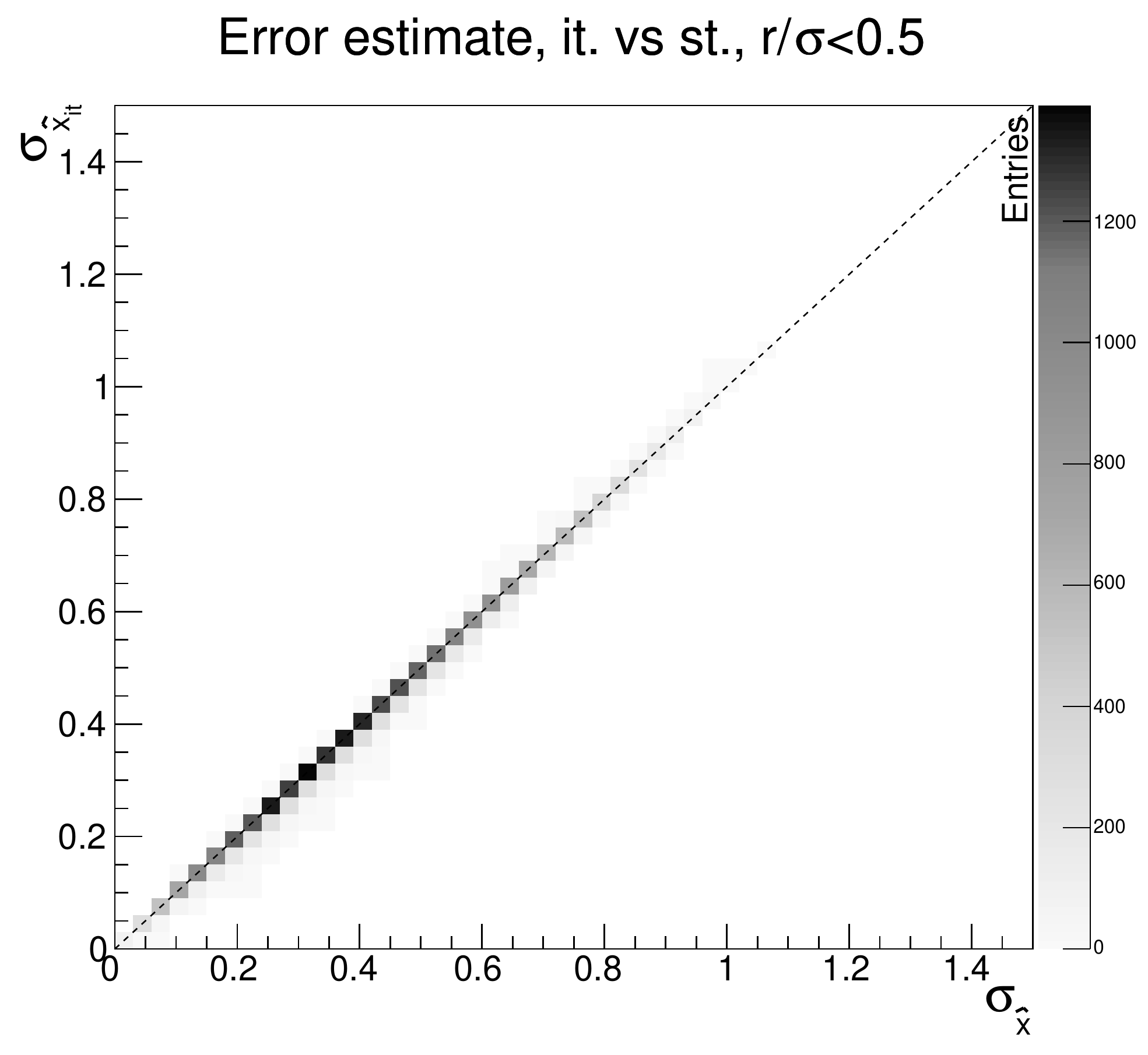}
 \caption{Distribution of the error estimates for the iterative versus standard the  BLUE methods 
 requiring $r_1, r_2$ (left) or $r_1/\sigma_1, r_2/\sigma_2$ (right) to be
 less than 0.9 (top), 0.75 (middle) or 0.5 (bottom).   }
\label{fig:rmsr}
\end{center}
\end{figure}
\begin{figure}[htbp]
 \begin{center}
 \includegraphics[width=0.49\textwidth]{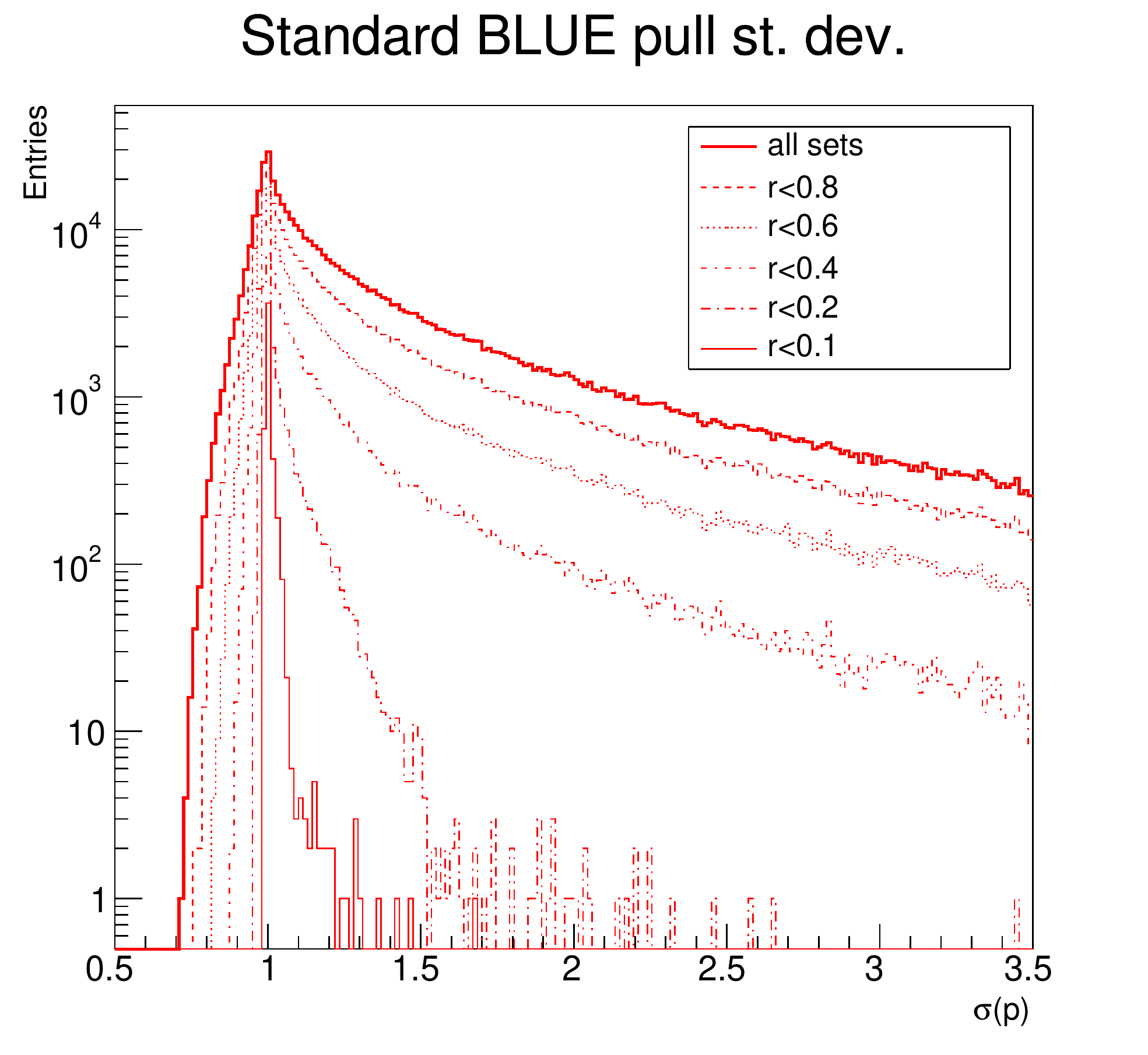}
 \includegraphics[width=0.49\textwidth]{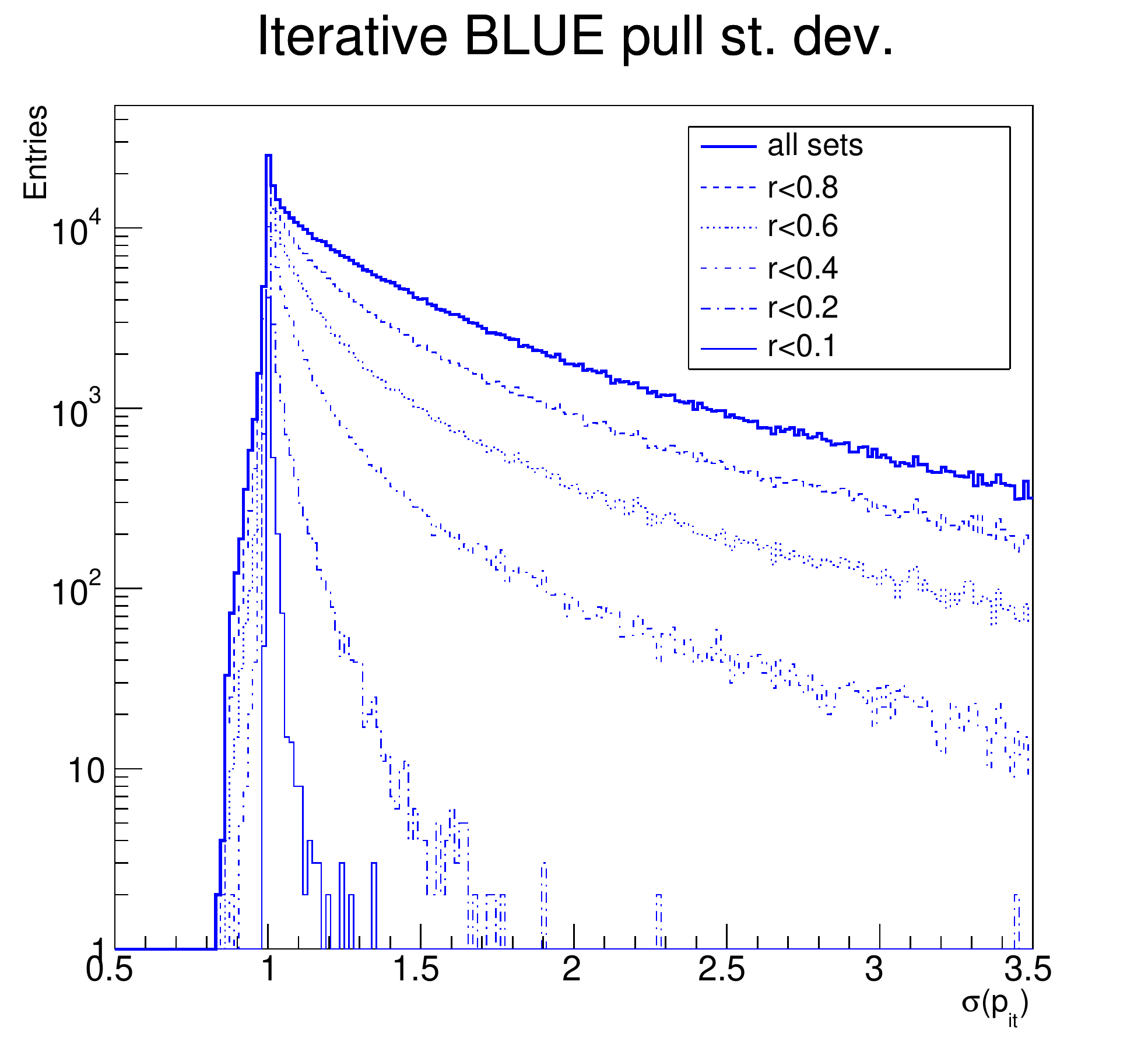}
 \includegraphics[width=0.49\textwidth]{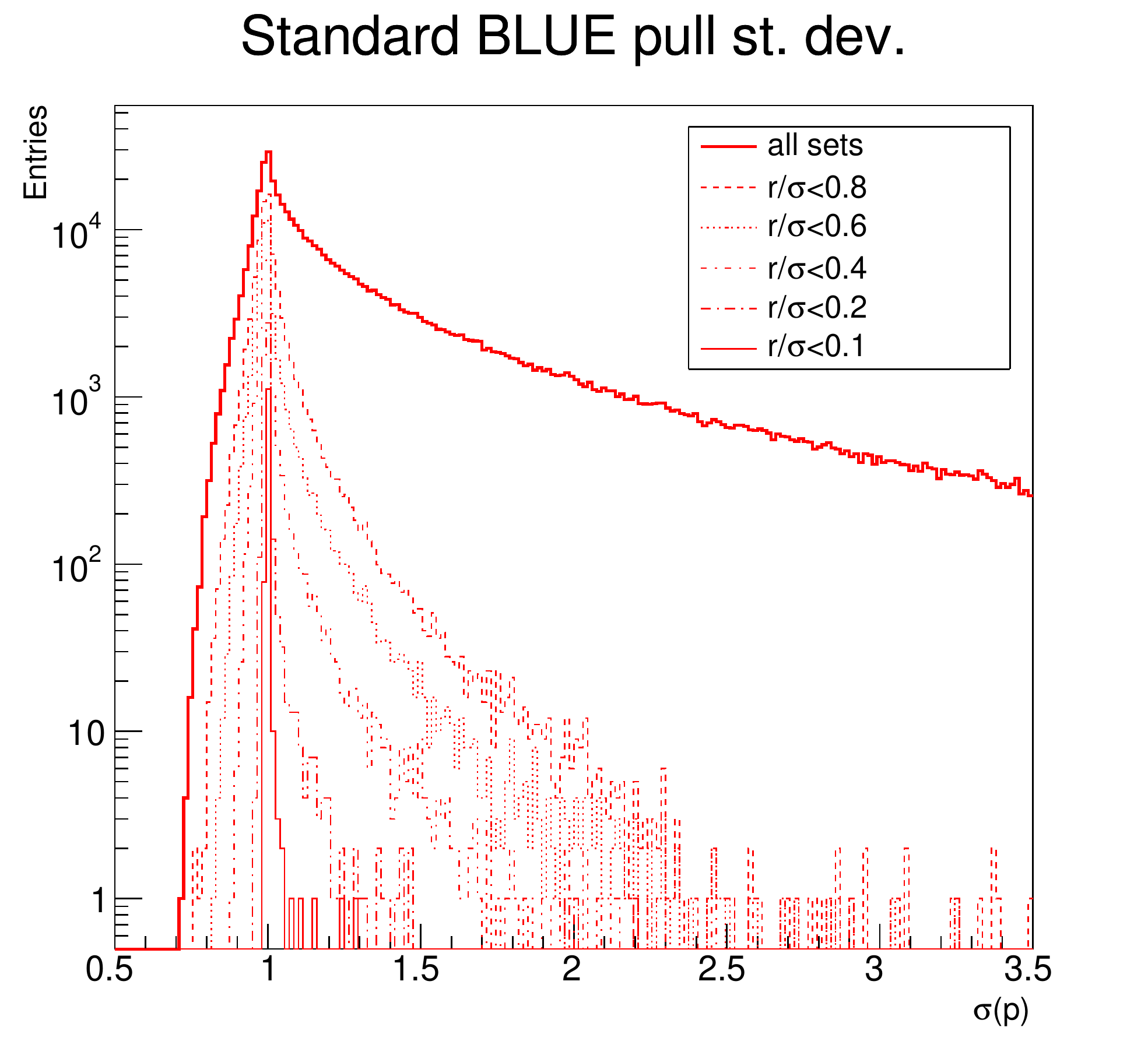}
 \includegraphics[width=0.49\textwidth]{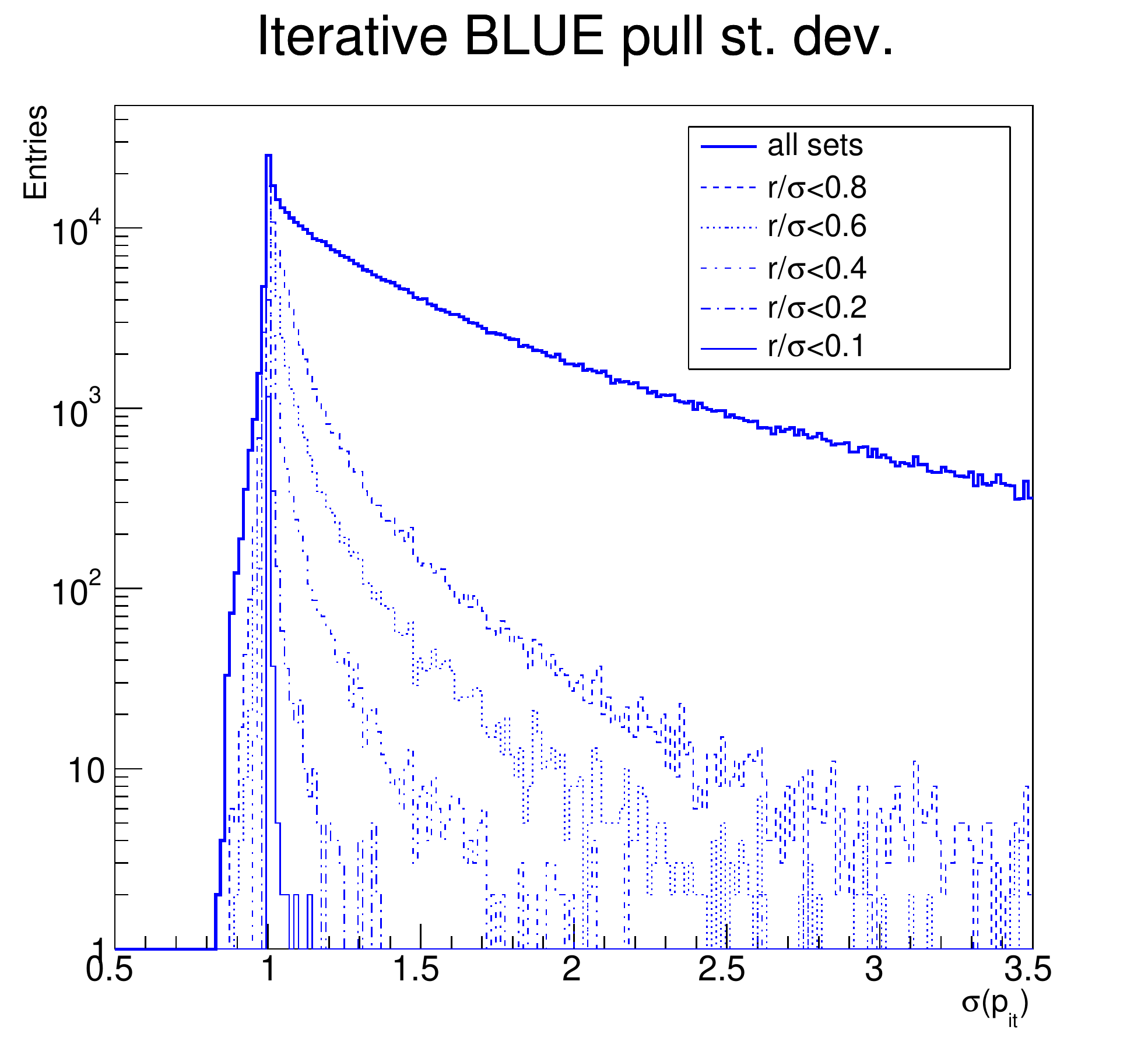}
 \caption{Average value of the pull standard deviation for the standard  (left) and iterative (right) BLUE estimates for different limits on $r$ (top) and on $r/\sigma$ (bottom).}
\label{fig:pullrms}
\end{center}
\end{figure}
\begin{figure}[htbp]
 \begin{center}
 \includegraphics[width=0.49\textwidth]{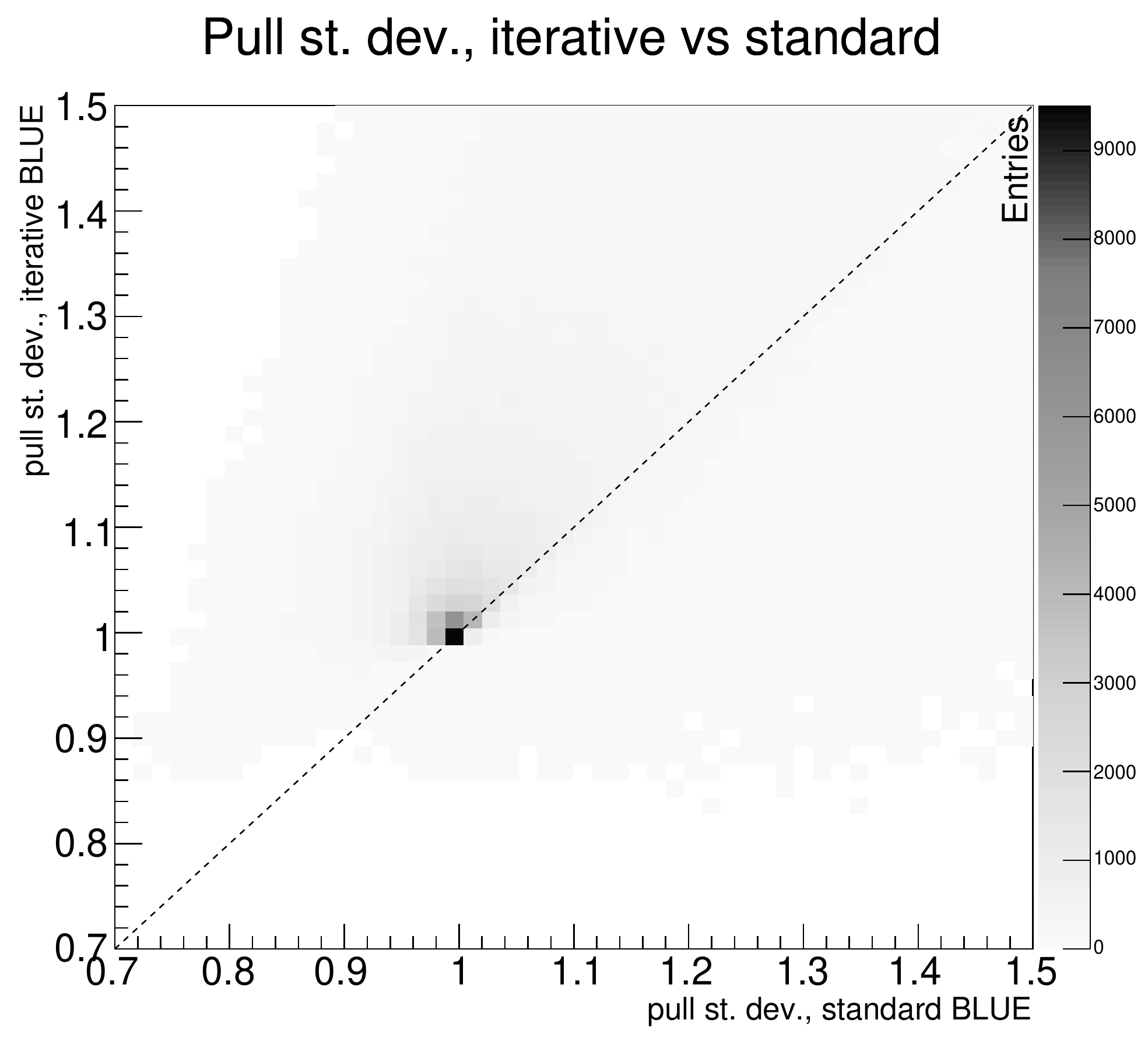}
 \includegraphics[width=0.49\textwidth]{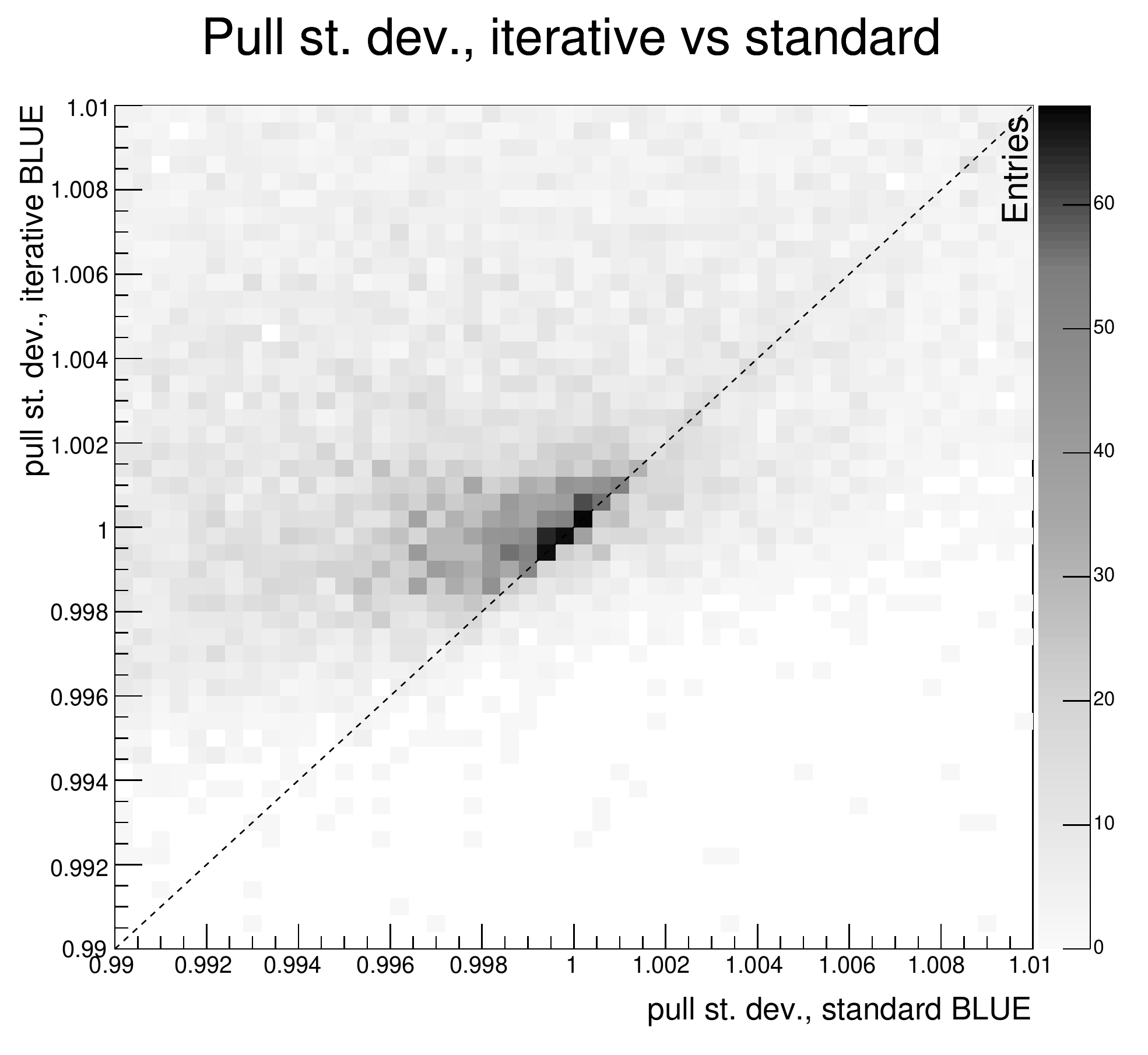}
 \caption{Average value of the pull standard deviation for the iterative vs standard BLUE estimates (right: zoom in the central region).}
\label{fig:pullrms2d}
\end{center}
\end{figure}
\begin{figure}[htpb]
 \begin{center}
 \includegraphics[width=0.49\textwidth]{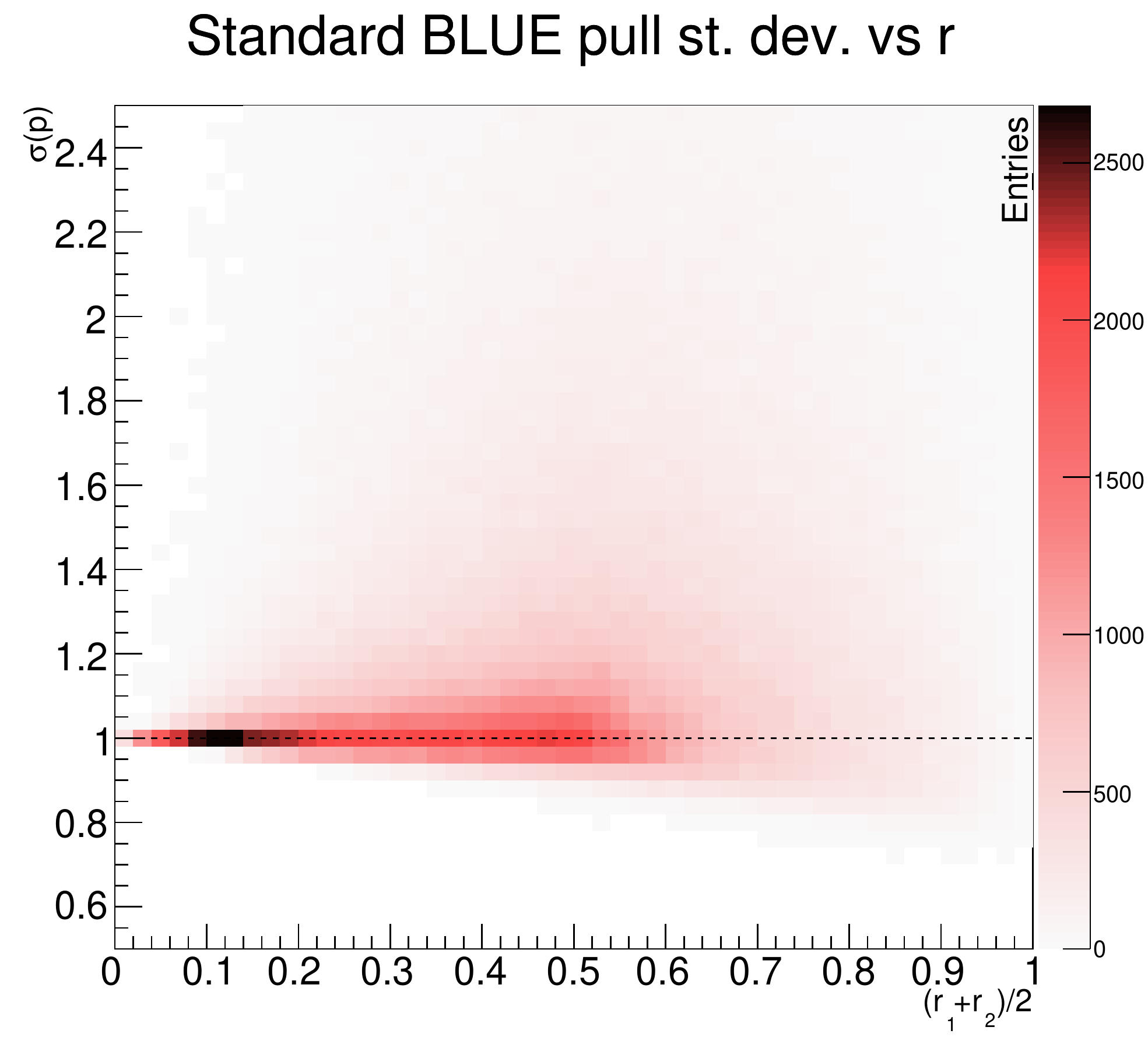}
 \includegraphics[width=0.49\textwidth]{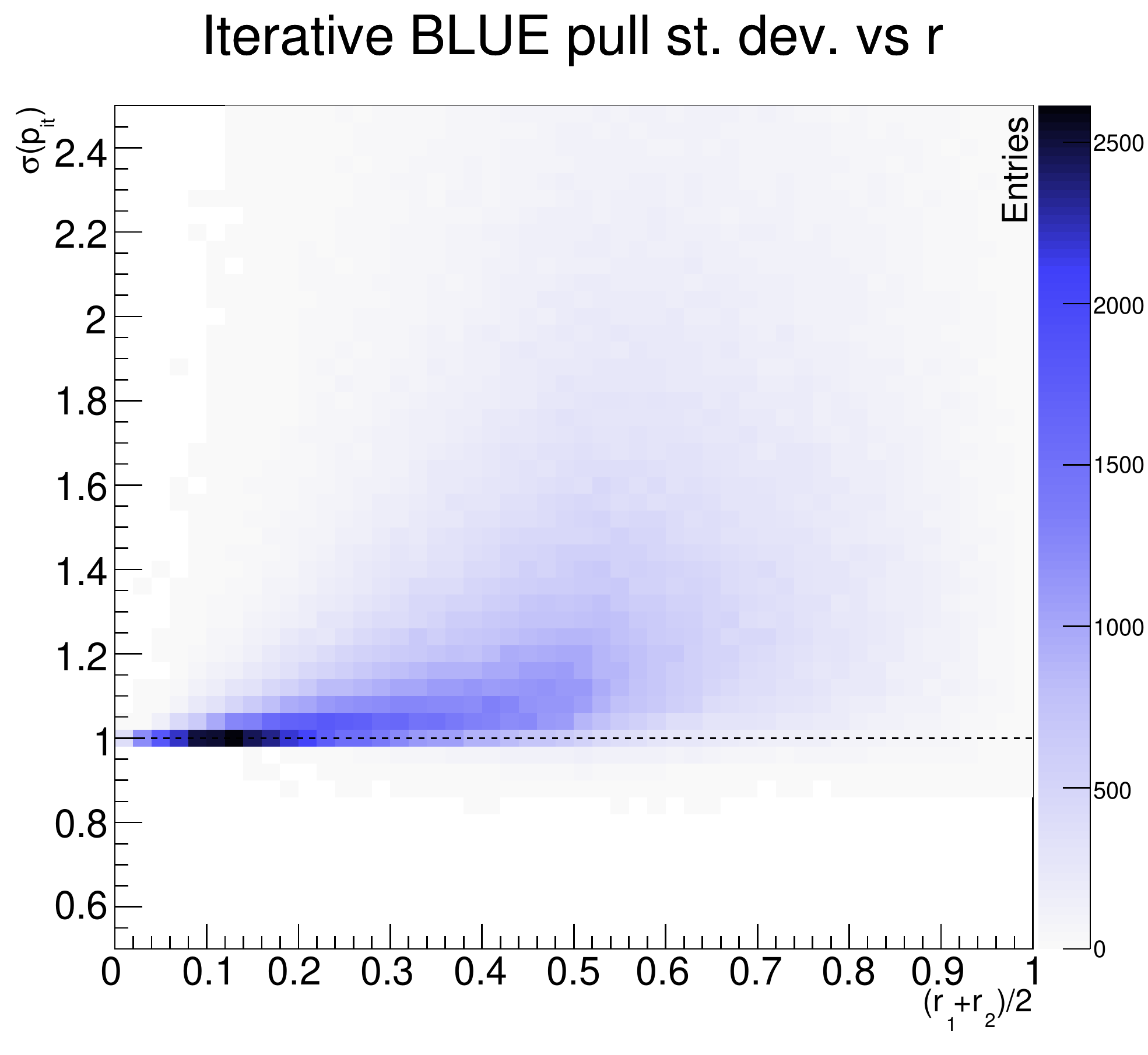}
  \includegraphics[width=0.49\textwidth]{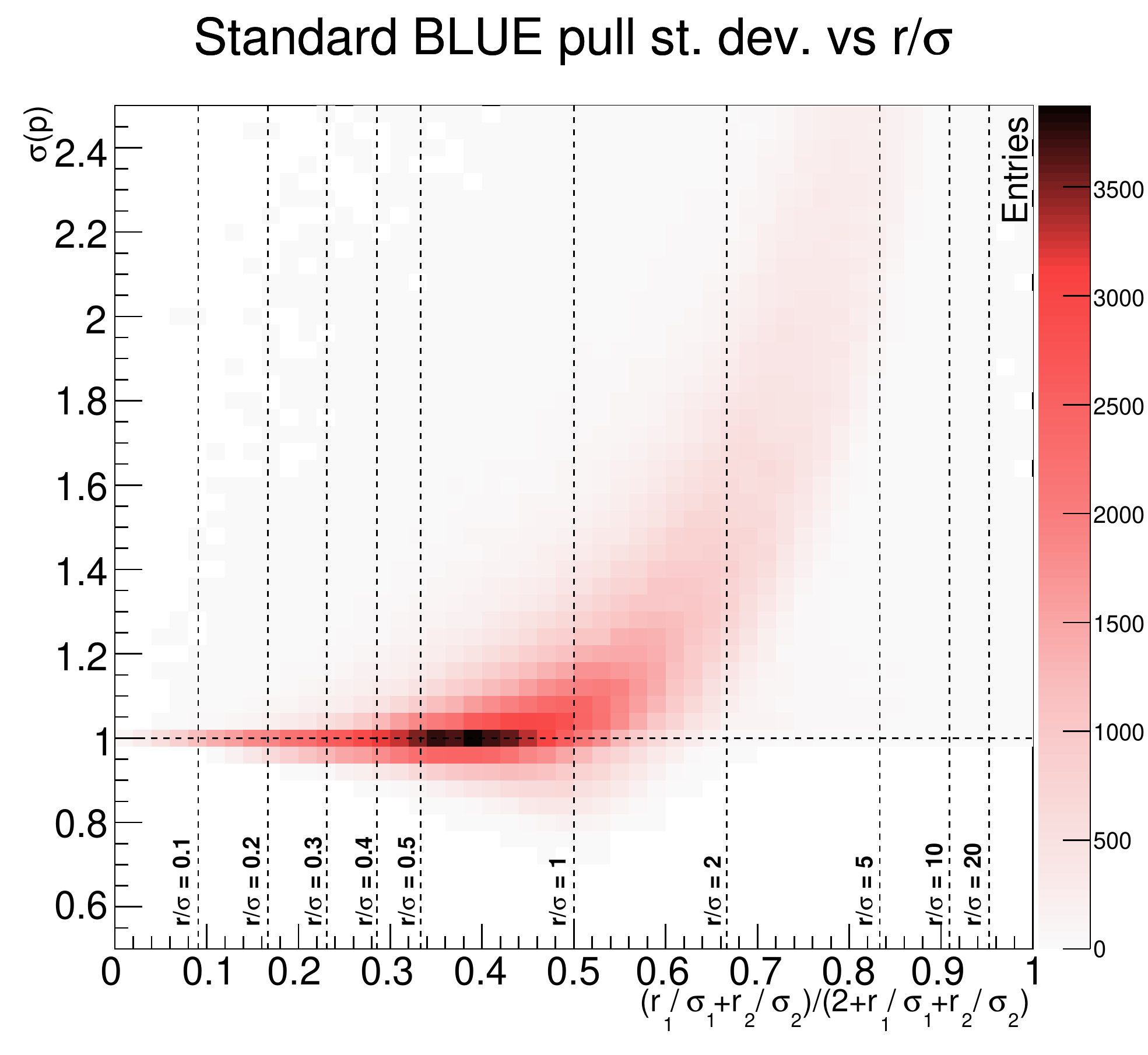}
 \includegraphics[width=0.49\textwidth]{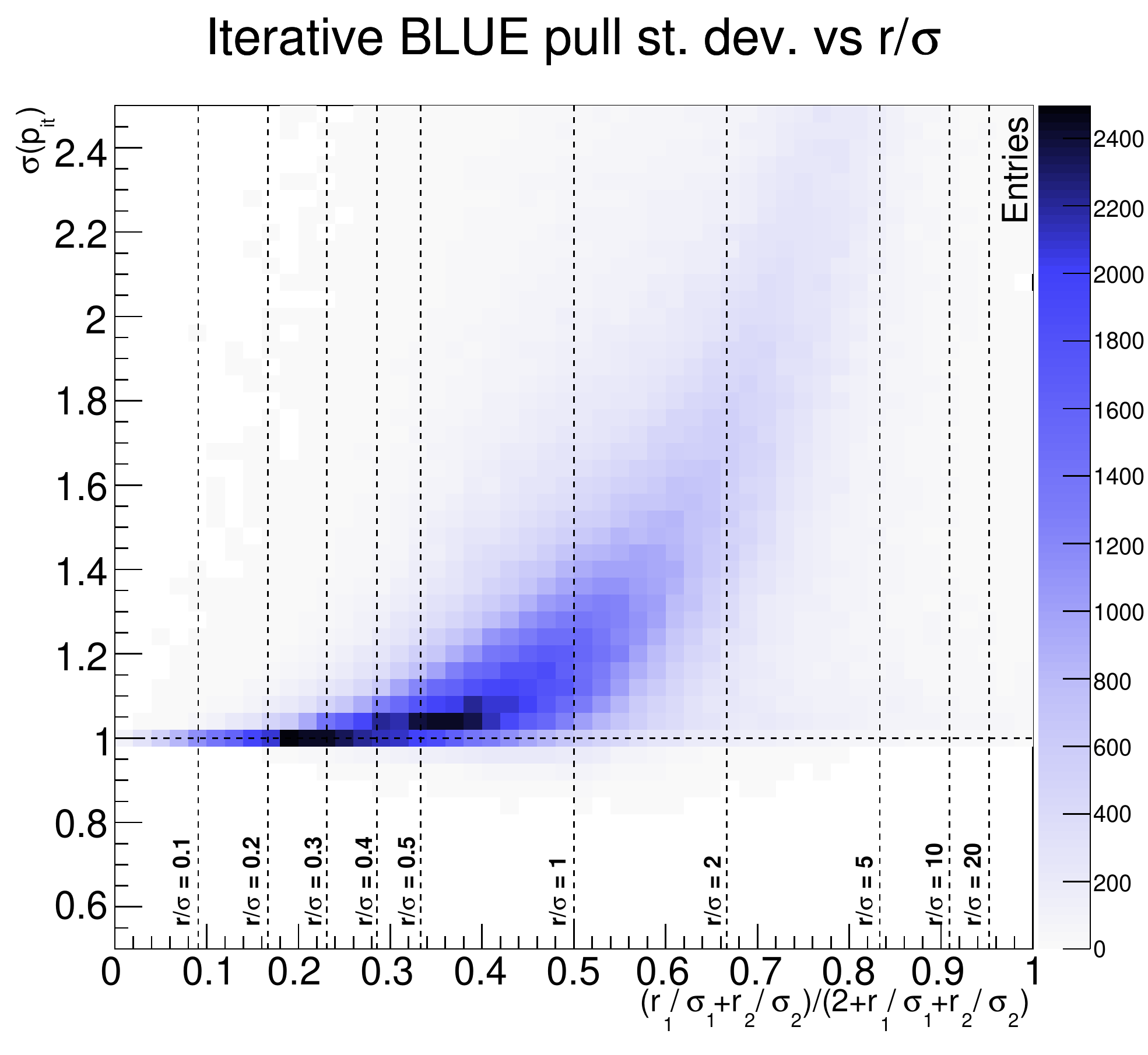}
 \caption{Distribution of the pull standard deviation measured with the standard (left) and iterative (right) BLUE method as a function
of $\frac{r_1+r_2}{2}$ (top) and 
$\frac{r_1/\sigma_1+r_2/\sigma_2}{2+r_1/\sigma_2+r_2/\sigma_2}$, which is a rescaling of $\frac{r_1/\sigma_1+r_2/\sigma_2}{2}$ (bottom).}
\label{fig:prms-rho}
\end{center}
\end{figure}
\begin{figure}[htpb]
 \begin{center}
 \includegraphics[width=0.49\textwidth]{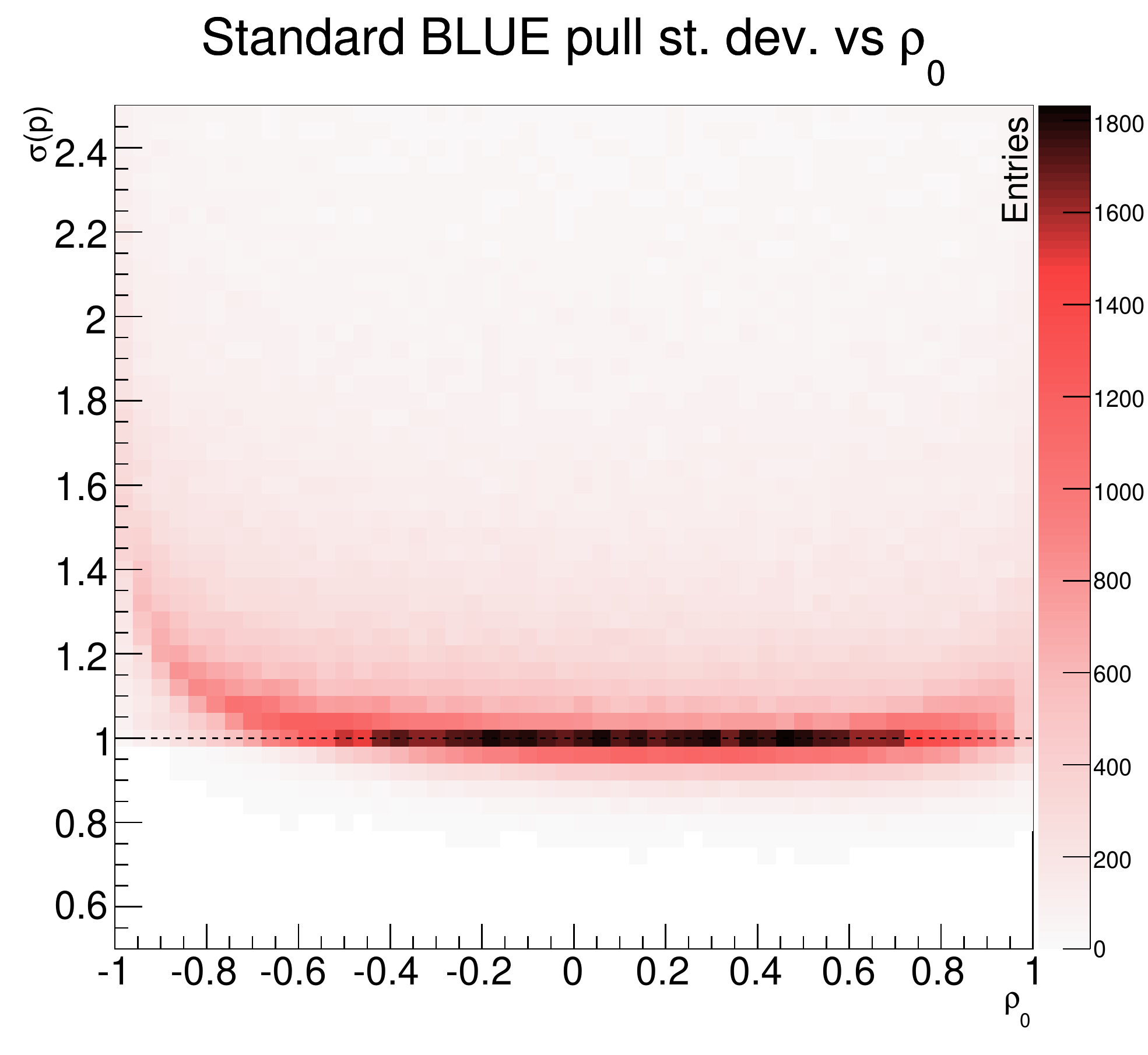}
 \includegraphics[width=0.49\textwidth]{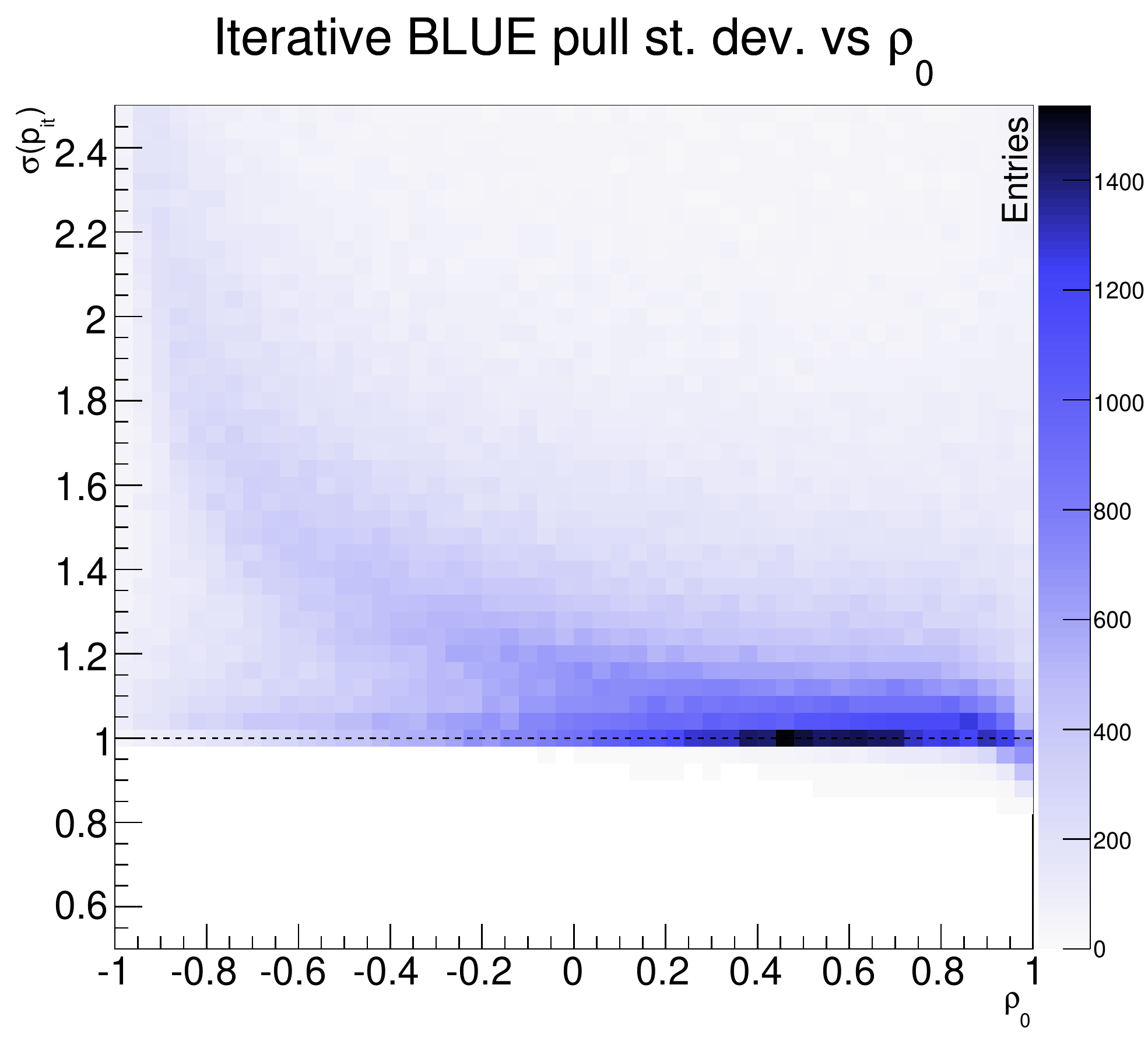}
 \includegraphics[width=0.49\textwidth]{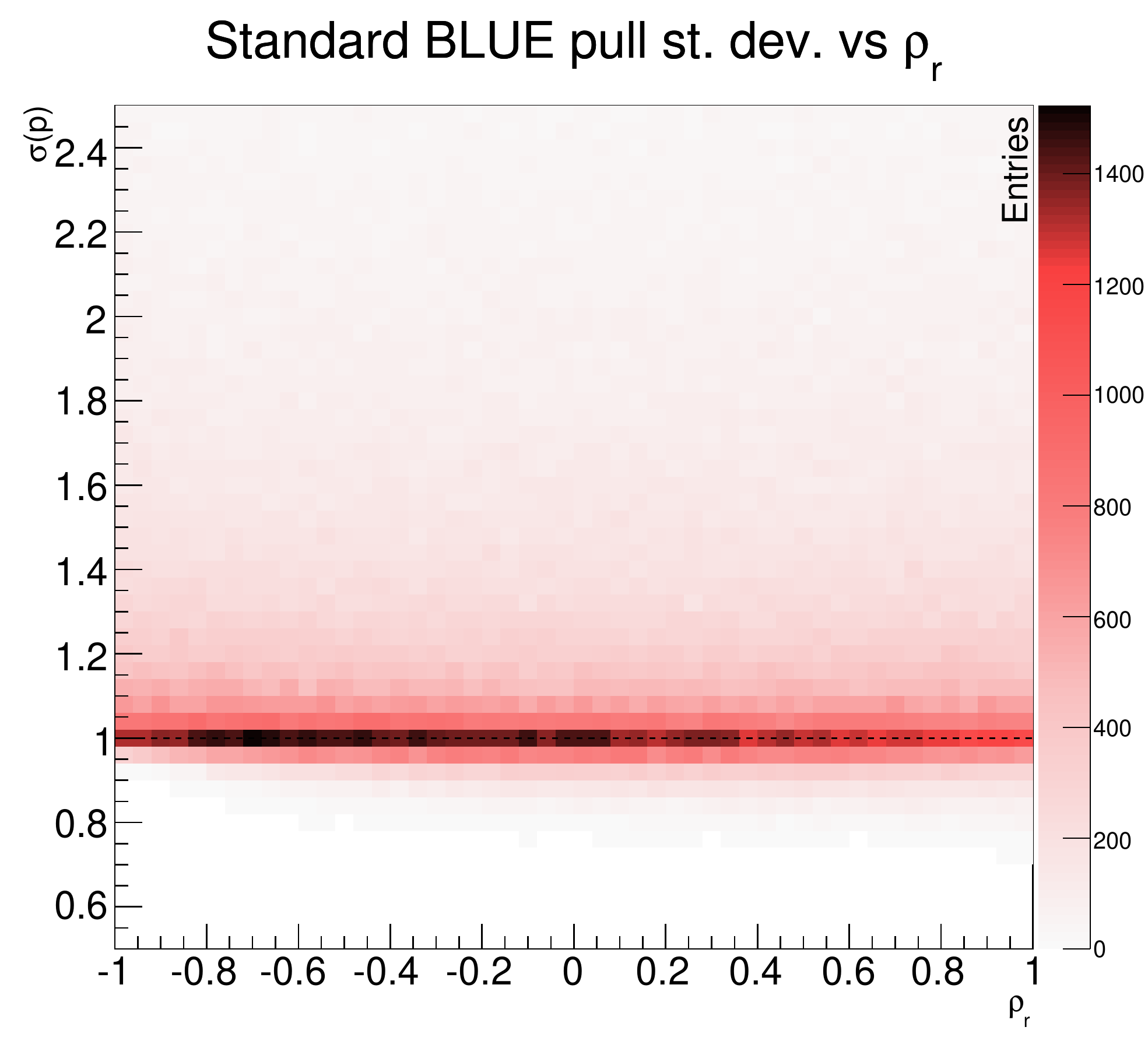}
 \includegraphics[width=0.49\textwidth]{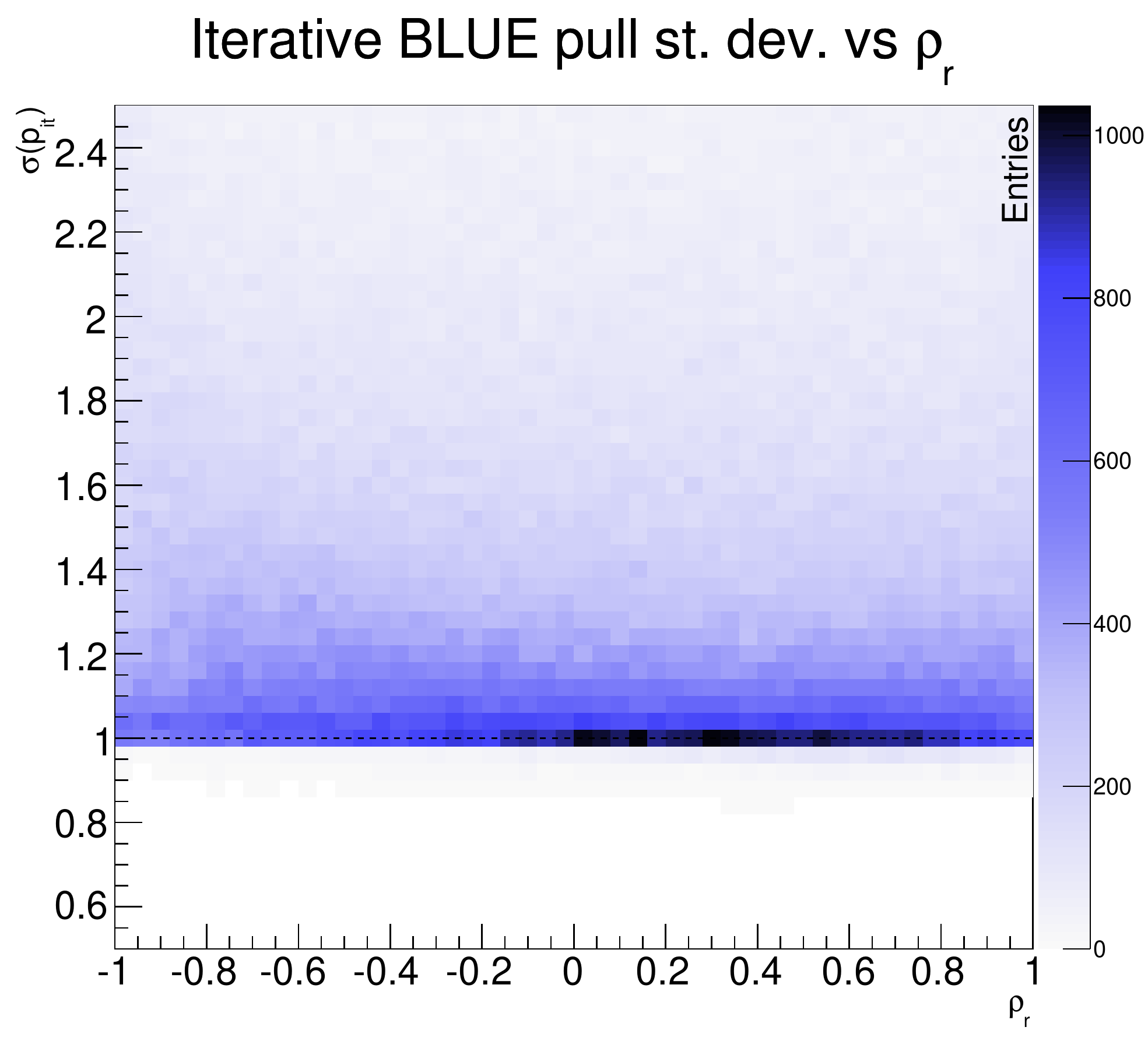}
 \caption{Distribution of the pull standard deviation measured with the standard (left) and iterative (right) BLUE method as a function of 
 $\rho_0$ (top) and $\rho_{\mathrm{r}}$ (bottom).}
\label{fig:prms-rhor}
\end{center}
\end{figure}

Figures~\ref{fig:rms} and~\ref{fig:rmsr} shows the distribution of the error estimates, using Eq.~(\ref{eq:blueerr}),
for the iterative estimator versus the standard estimator applying different requirements on $r$ or
$r/\sigma$. The uncertainty estimates of the two methods 
get closer as stronger requirements are applied on 
either  $r$ or for $r/\sigma$ and become
identical within few percent for $r< 0.5$ or for $r/\sigma<0.5$.
Anyway, the error estimates obtained from Eq.~(\ref{eq:blueerr}) and shown in Figs.~\ref{fig:rms} and~\ref{fig:rmsr}
may differ from the real standard deviation of the BLUE estimator, both in the standard and in the iterative case, when 
uncertainty estimates are used in place of the true ones. In order to test about the validity of those estimates,
Fig.~\ref{fig:pullrms} shows the distributions of the standard deviation obtained from the distributions of the estimators' pulls,
defined in Eq.~(\ref{eq:pull1}), (\ref{eq:pull2}) and Fig.~\ref{fig:pullrms2d} shows the distribution of the pull standard deviation for the iterative versus 
the standard BLUE estimates.
For an unbiased normally distributed estimator with correct error estimates, the pull distribution 
is a normal distribution with zero mean and standard deviation one. 
The mean of the pull distribution is a measure of the bias --which was studied separately-- and the standard deviation
 provides a test of the validity of the error estimate: if it is larger than one, it means that the error estimate 
is too small; if the standard deviation is smaller then one, the error estimate is too large.
For $r<0.1$ both the standard and the iterative estimators exhibit a pull standard deviation close to one within few percents
in most of the cases, but deviations may become significant as $r$ increases. The sensitivity of the 
pull standard deviation on $r/\sigma$ is smaller than on $r$, and for $r/\sigma<0.6$ the iterative estimator
has a pull standard deviation close to one within few percents, while deviations are larger for the standard estimator.
This dependency is also visible in Figs.~\ref{fig:prms-rho} and \ref{fig:prms-rhor}
that show the distribution of the pull standard deviation versus $r$, $r/\sigma$, $\rho_0$ and $\rho_{\mathrm{r}}$ for the two methods.
The pull standard deviation is mainly dependent on $r/\rho$ and, to a lesser extent, on $r$, while no evident correlation
with $\rho_0$ or $\rho_{\mathrm{r}}$ is visible from the plots.
In general, in most of the cases the error estimates of both the standard and the iterative methods tend to overestimate the 
uncertainty, but in some cases the standard method may also underestimate 
the uncertainty up to $20$--$30$\%, while this effect is reduced with the iterative method.


For physics application that report uncertainties with one significant digit, a relative uncertainty on the 
error estimate of 10\% may be sufficient. For those cases, the BLUE formula for the uncertainty may be
accurate enough in most of the cases if $r$ is below $\simeq 0.1$ or if $r/\sigma$ is below $\simeq 0.2$.
For larger relative uncertainties a dedicated study of the estimator's distribution (pull) may be a better
choice to determine the uncertainty in a more accurate way.

\section{Conclusions}
The application of the BLUE method and its iterative variant have been studied for the combination
of two measurements having uncertainty contributions that have a linear dependency on the measured value,
i.e.: in the case when relative uncertainty contributions are known.
The study was performed using a Monte Carlo simulation spanning a very
large range of possible values of the uncertainty contributions and their correlations. The study demonstrates
the possible presence of a significant bias in the application of the original ``standard'' BLUE method,
while in general the iterative application of the BLUE method significantly mitigates this bias.
In the cases having no extreme values of the uncertainty contributions that depend on the measured values, the bias of 
the standard BLUE method remains limited.

For the explored cases, both the standard and the iterative BLUE methods provide uncertainty estimates that
may differ from the true standard deviation of the estimator in some cases. The uncertainty estimate in the 
iterative method tends to provide underestimated errors which anyway agree within 10\% or better with the standard
deviation of the estimator in case the relative uncertainty contribution is smaller than 10\%, or
smaller than about 30\% of the remaining uncertainty contribution. For the other cases, in order to have a more
precise determination of the estimated uncertainty, it may be useful to determine the actual
variance using a dedicated study of the estimator's distribution for the specific case under investigation.

The present study covers the simplified case of the combination of two measurements.
A generalization to the combination of more measurements would be interesting, since
similar benefits of the iterative BLUE methods are expected.

\section{Acknowledgements}
I am very grateful to Pietro Santorelli who helped me explore the possibility to approach this
problem analytically.
I am grateful to Jochen Ott and Julien Donini for useful discussions and e-mail exchanges;
Jochen originally proposed the iterative application of the BLUE method for the
combination of single-top cross-section measurements in CMS.
I'd like to thank the TOPLHC working group, in particular Roberto Chierici for
useful editorial suggestions and the ATLAS colleagues for constructive
criticism about the application of the BLUE method.

\section*{References}

\bibliography{iter-blue}

\end{document}